\newif\ifsubmode
\submodefalse

\newif\ifprintfig
\printfigtrue

\ifsubmode
  \documentclass[referee]{aa}
  \usepackage{epsf}
  \usepackage{graphicx}
  \usepackage{aalongtable}
\else
  \documentclass{aa}
  \usepackage{epsf}
  \usepackage{graphicx}
  \usepackage{aalongtable}
\fi

\newcommand{\etal}{{et al.\ }}
\renewcommand{\deg}{^{\circ}}

\newcommand{\kms}{\>{\rm km}\,{\rm s}^{-1}}

\newcommand{\pc}{\>{\rm pc}}
\newcommand{\kpc}{\>{\rm kpc}}
\newcommand{\Mpc}{\>{\rm Mpc}}

\newcommand{\dpadg}{\Delta{\rm PA}_{\rm DG}}
\newcommand{\dpadj}{\Delta{\rm PA}_{\rm DJ}}

\begin{document}

\title{A dichotomy in the orientation of dust and radio jets in nearby low-power radio 
       galaxies\thanks{Based on observations with the NASA/ESA Hubble Space Telescope obtained at the Space Telescope Science Institute, which is operated by the Association of Universities for Research in Astronomy, Incorporated, under NASA contract NAS5-26555}
}
\titlerunning{A dichotomy in radio jet orientation}

\author{
G.A.~Verdoes Kleijn\inst{1} \and 
P.T.~de Zeeuw\inst{2}
}
\authorrunning{Verdoes Kleijn \& de Zeeuw}
\offprints{G.A.~Verdoes Kleijn}

\institute{European Southern Observatory, Karl-Schwarzschild-Strasse 2, 
                 85748, Garching bei M\"{u}nchen, Germany (gverdoes@eso.org)
                 \and
           Sterrewacht Leiden, Postbus 9513, 2300 RA Leiden, 
                 The Netherlands (dezeeuw@strw.leidenuniv.nl)
}

\date{Received 0000 0000 / Accepted 0000 0000}


\abstract{
We examine the properties of central dust in nearby quiescent and
active early-type galaxies. The active galaxies are low-power radio galaxies
with Fanaroff \& Riley Type I or I/II radio jets.  We focus on (a) the comparison
of the dust distributions in the active and quiescent galaxy samples
and (b) the relation between the radio jet and dust orientations.  Our
main observational conclusions are: (i) in line with previous
studies, the dust detection rate is higher in radio-jet galaxies than
in non radio-jet galaxies; (ii) radio galaxies contain a higher
fraction of regular dust 'ellipses' compared to quiescent galaxies
which contain more often irregular dust distributions, (iii) the
morphology, size and orientation of dust ellipses and lanes in
quiescent early-types and active early-types with kpc-scale radio jets
is very similar; (iv) dust ellipses are aligned with the major axis of
the galaxy, dust lanes do not show a preferred alignment except for
large ($>$kpc) dust lanes which are aligned with the minor axis of the
galaxy and (v) as projected on the sky, jets do not show a preferred
orientation relative to the galaxy major axis (and hence dust
ellipses), but jets are preferentially perpendicular to dust
lanes. 

We show that the dust ellipses are consistent with being nearly
circular thin disks viewed at random viewing angles. The
lanes are likely warped dust structures, which may be in the
process of settling down to become regular disks or are being perturbed by a non-gravitational force. We use the observed dust-jet orientations to constrain the
three-dimensional angle $\theta_{\rm DJ}$ between jet and dust. 
For dust-lane galaxies, the jet is approximately perpendicular to the
dust structure, while for dust-ellipse galaxies there is a much wider
distribution of $\theta_{\rm DJ}$.

We discuss two scenarios that could explain the dust/jet/galaxy orientation dichotomy.
If lanes are indeed settling, then the jet orientation apparently is roughly
aligned with the angular momentum of the dust before it settles. 
If lanes are perturbed by a jet-related force, it appears that it causes the dust to move out of its equilibrium plane in the galaxy into a plane which is perpendicular to the jet.    
}

\maketitle





\section{Introduction}
\label{s:intro}

Many early-type galaxies harbour dust and warm (i.e., T$\sim
10^{4-5}$K) gas in their central regions (e.g., Sadler \& Gerhard
\cite{Sad85}; Goudfrooij \etal \cite{Gou94}; van Dokkum \& Franx \cite{vDok95}). Evidence is
mounting that all early-type galaxies harbour a central supermassive
black hole (BH) (e.g., Kormendy \& Gebhardt \cite{Kor01}, and references
therein). The dust and gas form potential fuel for the BH to power an
active galactic nucleus (AGN). In the nearby Universe, many, perhaps
the majority, of nearby early-type galaxies display some sort of
nuclear activity, most often as a modest AGN (e.g., Ho \etal \cite{Ho97};
Kauffmann \etal \cite{Kau03}). Comparing the demography of the fuel reservoirs
in nearby active and quiescent galaxies can shed light on the
triggering and/or feeding mechanism of AGN. In a small fraction of the active galaxies the AGN is accompanied by
kpc-scale radio jets (e.g.,
Condon \& Broderick \cite{Con88}). Here we focus on those radio galaxies. The orientation and kinematics of the dust and gas
form a tracer of the stellar potential in addition to the stellar
photometry and kinematics (e.g., Merritt \& de Zeeuw \cite{Mer83}). This
requires that the interstellar material has settled in the
gravitational potential and is unperturbed by collisional forces.
The observed
distribution of relative orientations of the gas and dust, the jet,
and the stellar potential, can constrain the physical processes that
govern the orientation of dust and jet and perhaps the jet formation mechanism in
radio galaxies. 

Many ground-based studies have surveyed the global, kpc-scale, dust
content of nearby (distance $<100\Mpc$) quiescent and/or active
early-type galaxies (e.g., Sadler \& Gerhard \cite{Sad85}; Veron-Cetty \&
Veron \cite{Ver88}; Knapp \etal \cite{Kna89}; Roberts \etal \cite{Rob91}; Goudfrooij \etal
\cite{Gou94}). With the Hubble Space Telescope (HST) it has become possible to
detect routinely central dust distributions, also those with sizes below 1 kpc, in nearby
galaxies. Using HST/WFPC2 optical broad-band photometry, dust is
detected in $\sim 50\%$ of the nearby early-type galaxies (van Dokkum
\& Franx \cite{vDok95}; Tomita \etal \cite{Tom00}; Tran \etal \cite{Tra01}). This detection
rate increases to $\sim 90\%$ for nearby early-type galaxies with
radio jets (van Dokkum \& Franx \cite{vDok95}; Verdoes Kleijn \etal \cite{Ver99}). The
increased detection rate in radio galaxies suggests a causal
connection between extended central ISM distributions and the on-set
of nuclear activity and radio jet formation.

Early studies, using ground-based imaging, showed that jets in radio
galaxies are roughly perpendicular in the plane of the sky to kpc-scale dust
structures (e.g., Kotanyi \& Ekers \cite{Kot79}; M\"{o}llenhoff, Hummel
\& Bender \cite{Mol92}). No relation was found between galaxy and jet
orientation (e.g., Battistini \etal \cite{Bat80}; Birkinshaw \& Davies \cite{Bir85};
Sansom \etal \cite{San87}) apart from a tendency to avoid large $\ga 70\deg$
angles between galaxy minor axis and radio jet.  In studies of the sub
kpc-scale dust, based on HST imaging, the perpendicularity of jet and
dust in the plane of the sky was also found for radio galaxies (e.g., van
Dokkum \& Franx \cite{vDok95}; de Koff \etal \cite{dKof00}; de Ruiter \etal
\cite{dRui02}). Furthermore, evidence was reported that also the intrinsic,
i.e., three-dimensional, orientation of radio jets is roughly
perpendicular to the dust, using samples of radio
galaxies which contain dust disks (Capetti \& Celotti \cite{Cap99}; Sparks
\etal \cite{Spa00}). In contrast, Schmitt \etal (\cite{Sch02}) found in 
a sample of 20 radio galaxies with regular dust disks that the jets
are {\sl not} roughly perpendicular to the disks in three-dimensional
space.

We have studied a sample of nearby Fanaroff \& Riley (\cite{Fan74}) Type I
radio galaxies, the 'UGC~FR-I sample' (Xu \etal \cite{Xu00}; Verdoes Kleijn
\etal \cite{Ver99}, \cite{Ver02}; Noel-Storr \etal \cite{Noe03}). Verdoes Kleijn \etal (\cite{Ver99})
describe a dichotomy in apparent dust morphology of the UGC~FR-I
sample galaxies, which either contain regular dust structures with an
elliptical appearance ('ellipses'), or filamentary structures
('lanes'). Ellipses turn out to be aligned with the galaxy major axis
while the lanes show no relation to galaxy orientation.  Furthermore,
most lanes (but only some ellipses) are perpendicular to jets in the
sky-plane. Clearly, the apparently conflicting results on the dust-jet
orientation mentioned above might be connected to this
relation between dust orientation and morphology. We explore this relation further by inferring the intrinsic three-dimensional relative
orientations of jet, dust and stellar content of the host galaxy. With
the results of this analysis we hope to constrain the causal
connections which lead to the observed orientations between the following
three actors: (i) the gravitational potential, (ii) the central ISM
and (iii) the radio jet. We focus on the following questions. Are
the properties of the fuel reservoirs different in radio galaxies and
non-radio galaxies? Is the dust orientation governed by the
gravitational potential? Do jet/AGN related forces affect the dust
orientation?

This paper is organized as follows. In Section~\ref{s:samples} we
describe the selection and properties of the radio and non-radio
galaxy samples. In
Sections~\ref{s:dustproperties} and \ref{s:dustcomparison} we compare
the dust properties in radio and non-radio galaxies as derived from
HST/WFPC2 imaging. In Section~\ref{s:dustjet} we describe the relative
orientations of jets, dust and the galaxy in the plane of the sky. In
Sections~\ref{s:ellipsecircular} and \ref{s:ellipsejetintrinsic} we
determine the constraints placed by the projected orientations on the
intrinsic relative orientations in three dimensions for galaxies with
dust classified as {\sl ellipses}. We consider the dust {\sl lane}
galaxies in Section~\ref{s:lanejetintrinsic}. In
Section~\ref{s:discussion} we discuss the possible origins for the
dichotomy in intrinsic orientation of dust and jets found in Sections~\ref{s:ellipsejetintrinsic} and~\ref{s:lanejetintrinsic}.  We summarize our
results in Section~\ref{s:conclusions}.

Throughout the paper we use a Hubble constant $H_0$= 75
kms$^{-1}$Mpc$^{-1}$.

\section{Samples and Data}
\label{s:samples}

We use a complete sample of 21 radio (i.e., 'active') galaxies, the
UGC~FR-I sample, and a sample of 52 non-radio (i.e., 'quiescent')
galaxies, the UGC~non-FR-I sample, to statistically compare dust
distributions. We also consider a larger 'FR sample', which
contains 47 FR-I and FR-I/II radio galaxies in total, to improve the constraints on the
three-dimensional relative orientation of jets, dust and stellar
hosts.  We rely mainly on HST/WFPC2 imaging. The exceptions are 3C
402N, for which we use HST/FOC imaging, and NGC 5128 for which we
employ VLT/VIMOS imaging which provides a full view of the kpc-scale
dust distribution in this nearby galaxy.

\subsection{The Radio Galaxy Samples}
\label{s:frisample}

The {\bf UGC~FR-I sample} is a complete sample of nearby radio
galaxies from the UGC~catalogue (Nilson \etal \cite{Nil73}). It contains all 21
nearby ($v < 7000\ \rm{km~s^{-1}}$), elliptical or S0 galaxies
in the declination range $-5^\circ < \delta < 70^\circ$ in the
UGC~catalogue (magnitude $m_B < 14\fm 6$ and angular size $\theta_p >
1\farcm 0$) that are extended radio-loud sources, defined as larger
than 10\arcsec\ at $3\sigma$ on VLA A-Array maps and brighter than
150~mJy from single-dish flux-density measurements at 1400 MHz. These
galaxies have jets classified as Fanaroff \& Riley (\cite{Fan74}) Type I
(FR-I).  HST imaging for the UGC~FR-I sample is presented in Verdoes
Kleijn \etal (\cite{Ver99}, \cite{Ver02}). Xu \etal (\cite{Xu00}) and Noel-Storr
\etal (\cite{Noe03}) report VLBA radio imaging and HST optical spectroscopy,
respectively, for the sample.\looseness=-2

We searched the literature for additional galaxies with FR-I or FR-I/II radio
sources which show dust on HST imaging data. This resulted in 26
galaxies from the 3CR and B2 radio galaxy catalogues and one 4C
galaxy. We refer to this extended sample as the {\bf FR sample}, and
list the general properties of the 47 FR-I galaxies in
Table~\ref{t:fr1}. The HST imaging data for 3CR and B2 galaxies are
described in de Koff \etal (\cite{dKof96}), Martel \etal (\cite{Mar99}), Capetti
\etal (\cite{Cap00}) and Sparks \etal (\cite{Spa00}) while the imaging for 4C-03.43 is described in Schmitt
\etal (\cite{Sch02}).

\subsection{The Non-Radio Galaxy Sample}
\label{s:nonfrisample}

\begin{figure*}[t]
\begin{centering}
\resizebox{1.0\hsize}{!}{\includegraphics{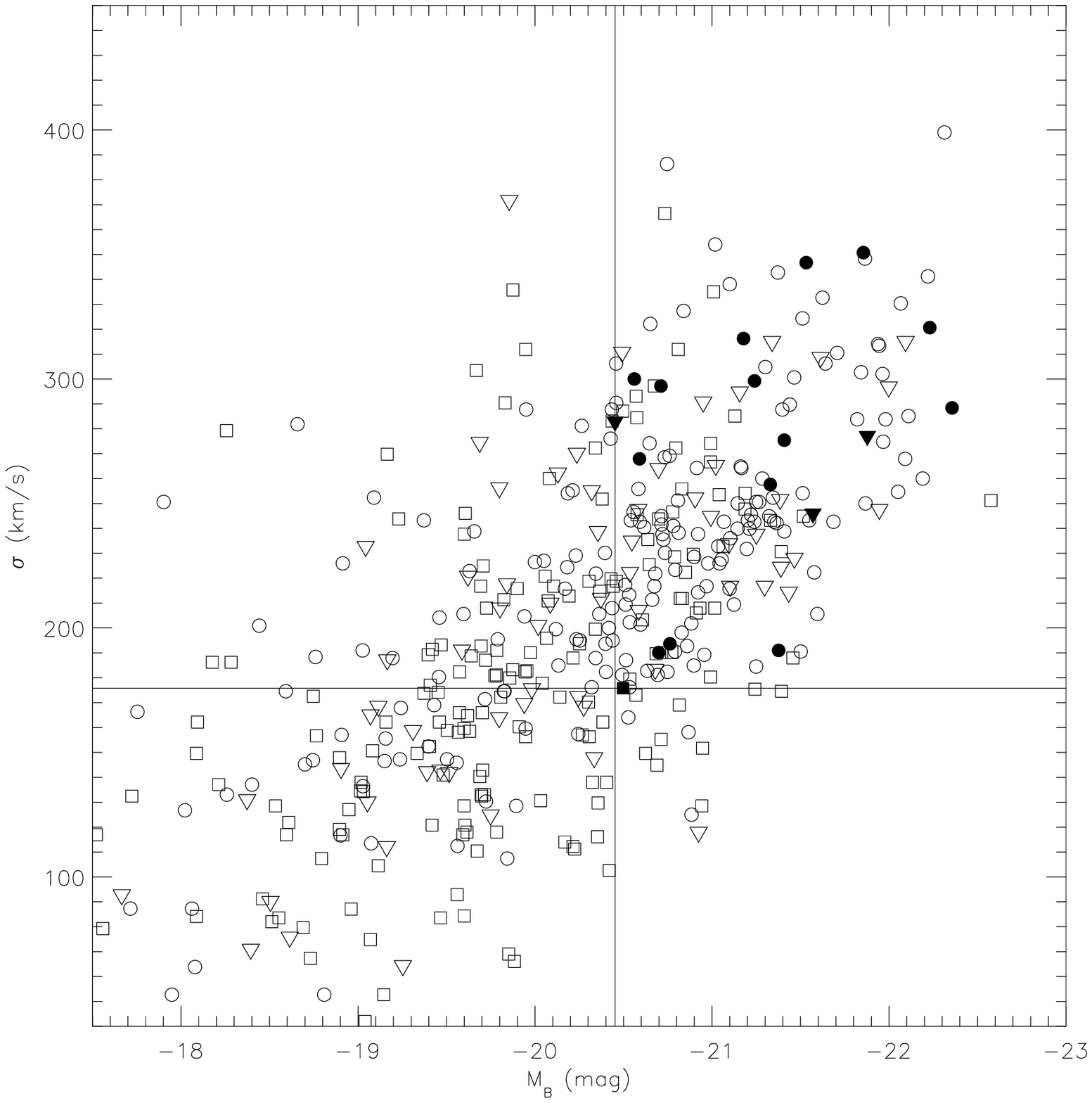}}
\caption{Central stellar velocity dispersion $\sigma$ 
  versus total absolute blue magnitude $M_{\rm B}$ for early-type
  galaxies in the Uppsala General Catalogue at $v<7000\kms$. Circles
  indicate ellipticals, triangles are E/S0 transition objects and
  squares denote S0s. The filled symbols indicate the 18 radio galaxies
  of the UGC~FR-I sample for which a $\sigma$ has been published.  The horizontal and vertical solid lines
  mark the minimum $\sigma$ and galaxy luminosity observed for the
  radio galaxies. Magnitudes, dispersions and morphologies are taken
  from the LEDA catalogue.}
\label{f:ugc}
\end{centering}
\end{figure*}

Figure~\ref{f:ugc} shows the location of the UGC~FR-I sample members
among the galaxies of the UGC in the plane of central stellar velocity
dispersion versus host absolute magnitude. The data are taken from the
LEDA catalogue\footnote{Lyon-Meudon Extragalactic Database
(http://leda.univ-lyon1.fr)} except for the velocity dispersion of 3C
66B and 3C 449 (Balcells \etal \cite{Bal95}) and NGC 4335 (Verdoes Kleijn
\etal \cite{Ver02}). The location of the UGC~FR-I galaxies confirms the well-known 
fact that radio galaxies are bright early-type galaxies.  The figure
also shows that the FR-I galaxies are distributed fairly uniformly,
relative to the general galaxy distribution, in the upper-right corner
of large host magnitude and velocity dispersion. The
morphologies of the UGC~FR-I galaxies are not a random selection from
the morphology distribution in this upper-right corner.
Table~\ref{t:samples} shows that the UGC~FR-I galaxies are
predominantly ellipticals and E/S0 transition objects and rarely pure
S0s.

\begin{figure*}[t]
\begin{centering}
\resizebox{1.0\hsize}{!}{\includegraphics{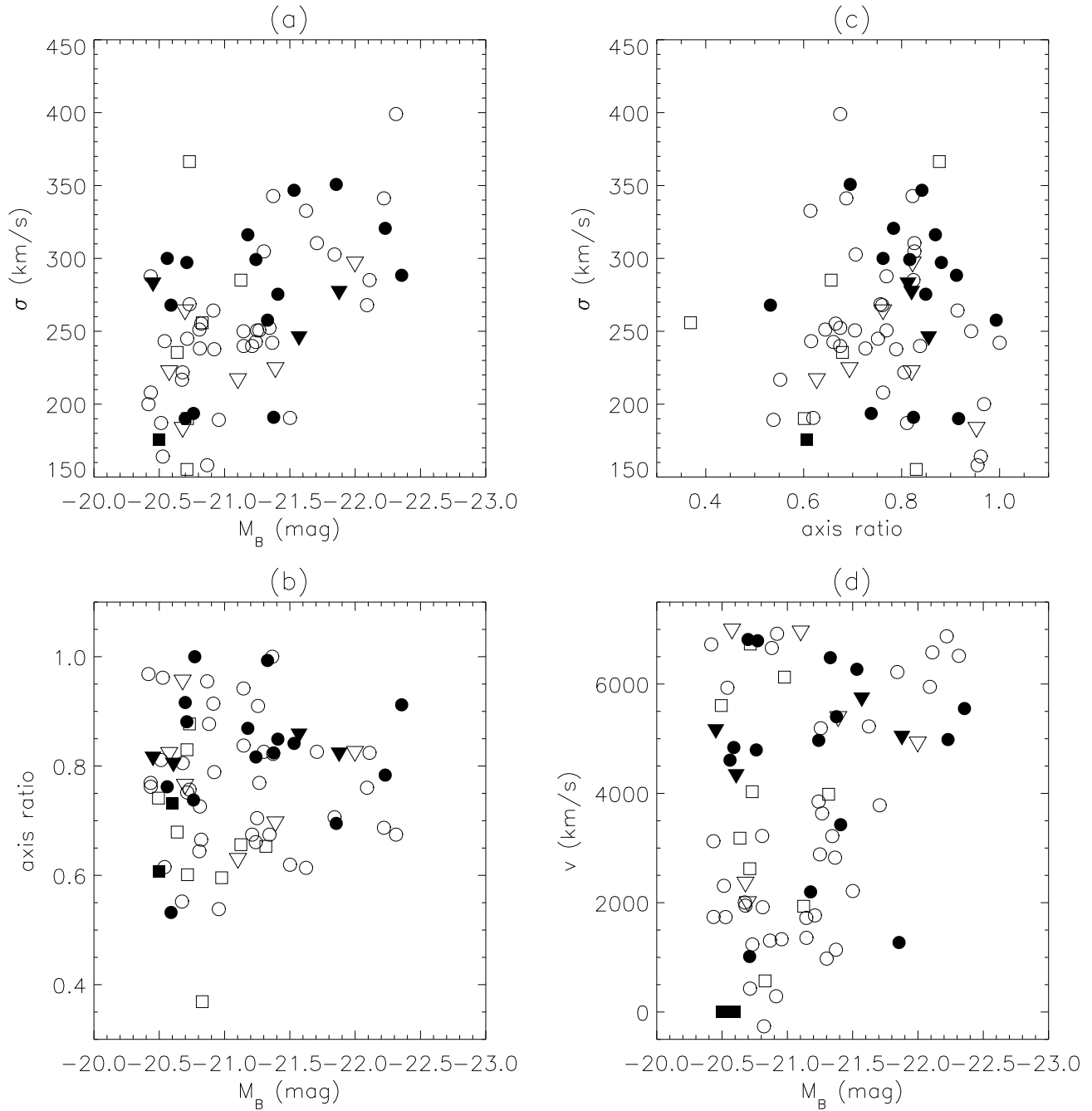}}
\caption{Central stellar velocity dispersion $\sigma$, host absolute blue magnitude $M_{\rm B}$, isophotal
axis ratio (at a blue surface brightness of 25 mag/arcsec$^{-2}$) and recession velocity $V$ for the UGC non-
FR-I sample of early-type galaxies (open symbols) and  the UGC~FR-I radio galaxies (filled symbols). Circles
indicate ellipticals, triangles E/S0 are transition objects and squares denote S0s.}
\label{f:ugcnormalsample}
\end{centering}
\end{figure*}

To carry out a statistical comparison of the dust distributions in
early-type galaxies with and without radio jets, we need samples of
both classes of galaxies with similar global properties of their
stellar host and similar HST imaging of the central regions so that we
can perform an identical dust analysis. Ideally, one would like to
perform the dust analysis for a well-defined sample of quiescent
galaxies in the UGC catalogue from which also the radio galaxies were
selected. Tran \etal (\cite{Tra01}) analysed dust features in a
representative mostly quiescent sample of nearby ($v<3400\kms$)
early-type galaxies using HST/WFPC2 observations. Unfortunately, this sample
contains galaxies with a typical host magnitude $\sim$2 magnitudes
fainter than in the UGC~FR-I sample. It also has a fraction of
lenticular galaxies which is about four times higher. We therefore
cross-correlated the HST/- WFPC2 archive with the UGC catalogue to
select all early-type UGC galaxies at $v<7000\kms$ which have a
similar absolute magnitude as the radio galaxies (i.e., $M_{\rm B} <
-20.4^m$), but which are not part of the UGC~FR-I sample. The
resulting sample of 52 galaxies is listed in Table~\ref{t:ugcnormal}
and its general properties are shown in
Figure~\ref{f:ugcnormalsample}. Some of the non-FR-I galaxies host
low-level activity in their nuclei, e.g., nuclear radio emission and
optical line-emission. For clarity we therefore refer to this sample
as the {\bf UGC non-FR-I sample} instead of, for example, the
quiescent sample. However, none of the UGC non-FR-I galaxies have jets
on scales of tens of kiloparsec or larger. The UGC non-FR-I sample
turns out to match the UGC~FR-I sample in velocity dispersion,
recession velocity and morphologies (Figure~\ref{f:ugcnormalsample} and Table~\ref{t:samples}).

The UGC~FR-I and UGC~non-FR-I samples also match in imaging depth and
filter coverage. The entire UGC~FR-I sample was observed in $V$- and
$I$-band.  Table~\ref{t:ugcnormal} shows that 88\% of the galaxies has
$V$ and/or $I$-band observations, 54\% has $V$- and $I$-band
observations and 12\% has only $R$-band observations. Thus the filter
coverage is similar for both samples. The many targets with both $V$-
and $I$-band observations allow us to verify that the dust detection
rate and the derived dust properties do not depend on the filter
used. For the UGC~FR-I sample the exposure times for both $V$- and
$I$-band images are typically 460s (increasing up to 1400s in a small
subset of the observations). Table~\ref{t:ugcnormal} shows that the
exposure times for the UGC~non-FR-I galaxies are similar.

\section{Dust Properties and Position Angle Measurements}
\label{s:dustproperties}

It is likely that every early-type galaxy contains dust due to mass loss from evolved stars. In addition, it can contain
externally acquired dust from tidal interactions or accretion of
galaxy satellites. Whether the dust obscuration is detected in optical
observations depends on various factors, including sensitivity of the
observations, dust morphology (e.g., smooth or clumpy) and orientation
of dust with respect to the observer (e.g., face-on or edge-on). To
ensure a homogeneous dust classification and measurement of dust
properties for all samples we re-analysed the published HST imaging
using the HST archive. We define `dust' as a localized depression in
the stellar surface brightness. Thus we are insensitive to a global
smooth dust component (e.g., following the stellar distribution) if it
were present in early-type galaxies. Our approach is similar to that
used in earlier HST imaging studies (e.g., van Dokkum \& Franx \cite{vDok95};
Verdoes Kleijn \etal \cite{Ver99}; Tran \etal \cite{Tra01}).

The appearance of the dust structures in both radio and non-radio
galaxies varies from regular elliptical shapes, evoking the idea of
inclined disks, to highly irregular structures, suggesting unsettled
dust. We use the classification scheme for the UGC~FR-I sample
described in Verdoes Kleijn \etal (\cite{Ver99}). It divides the morphology of
the dust structure as projected on the plane of the sky into three
bins. Class one, a 'dust lane', is a filamentary structure which
passes through the centre and which is sufficiently regular to assign
an orientation to it.  Class two, a 'dust ellipse', is a dust
structure with a circumference that resembles an ellipse. In some
galaxies it is not possible to determine unambiguously if the dust has
an ellipse- or a lane-like appearance. This is often due to the
faintness and/or rather small angular size of the dust
distribution. We classify these galaxies as 'intermediate' (see
Tables~\ref{t:fr1} and~\ref{t:dustugcnormal}). Class three, 'irregular
dust', is extended clumpy and/or filamentary dust which either is too
irregular to assign an orientation or does not pass through the
nucleus and hence is not classified as a lane. Some galaxies which
host a dust ellipse or lane also harbour a more extended, class three,
irregular dust distribution (see Tables~\ref{t:fr1} and
~\ref{t:dustugcnormal}). We discuss in
Sections~\ref{s:ellipsecircular} and~\ref{s:discussion} how the dust
classification scheme, based on appearance, depends on viewing angle
towards the dust distribution.

We define the size of the dust as the largest linear extent of the
dust feature. This is the dust major axis for dust ellipses. We
measure the ellipticity $\epsilon$ of dust ellipses using their
circumference.  Dust classes one and two, the lanes and ellipses, are
regular and extended enough to define also a position angle PA$_{\rm
D}$. For lanes it is the PA of the filament, taken as close
as possible to the nucleus if any bending is present. For ellipses
it is the PA of the dust ellipse major axis. For close-to-round
ellipses ($\epsilon<0.1$) we cannot determine a reliable PA of the major axis. The dust classification and dust properties for the FR-I and UGC
non-FR-I samples are listed in Tables~\ref{t:fr1} and
\ref{t:dustugcnormal}, respectively. As a cross-check we compared our
measurements of dust position angle and ellipticities with
measurements available in the literature for various galaxies
(Table~\ref{t:comparedust}). Ellipticities differ by $\la 0.05$ and PA$_{\rm D}$ differ by $\la10\deg$.

We define the galaxy orientation as the position angle of the stellar
isophotal major axis just outside the radius of the main dust distribution.
For the UGC non-FR-I, the B2 and some 3CR galaxies, we measured the PA ourselves
from isophotal fitting to the WFPC2 images, masking dusty regions. The PA was taken from the
literature for the remaining radio galaxies (see Table~\ref{t:fr1} for
references).
  
For the radio galaxies we also require the position angle PA$_{\rm J}$
of the jet axis close to the galaxy centre. For the UGC~FR-I sample,
we use the PA$_{\rm J}$ as determined by Xu \etal (\cite{Xu00}). These PAs
are obtained from VLBA measurements, or, if those are not available,
from the larger-scale VLA measurements. In cases where both VLA and
VLBA measurements are available, the difference in PA$_{\rm J}$ is
$4\deg$ on average and always $\le 14\deg$. For the other radio
galaxies the PA$_{\rm J}$ was taken from various studies (see
Table~\ref{t:fr1} for references). Given the difference in PA$_{\rm
J}$ between VLA and VLBA measurements and the published values for
other jets, we estimate a typical error of $8\deg$ for all PA$_{\rm
J}$.\looseness=-2

We are interested in the relative orientation of host, dust and jet if
present. Thus we list in Tables~\ref{t:fr1} and~\ref{t:dustugcnormal}
the position angle difference $\Delta$PA$_{\rm DG}$ between dust and
galaxy axis and $\Delta$PA$_{\rm DJ}$ between dust and jet axis, both
defined in the range $[0\deg,90\deg]$.

\section{Comparison of Dust Properties} 
\label{s:dustcomparison}

\begin{figure*}[t]
\begin{centering}
\resizebox{0.5\hsize}{!}{\includegraphics{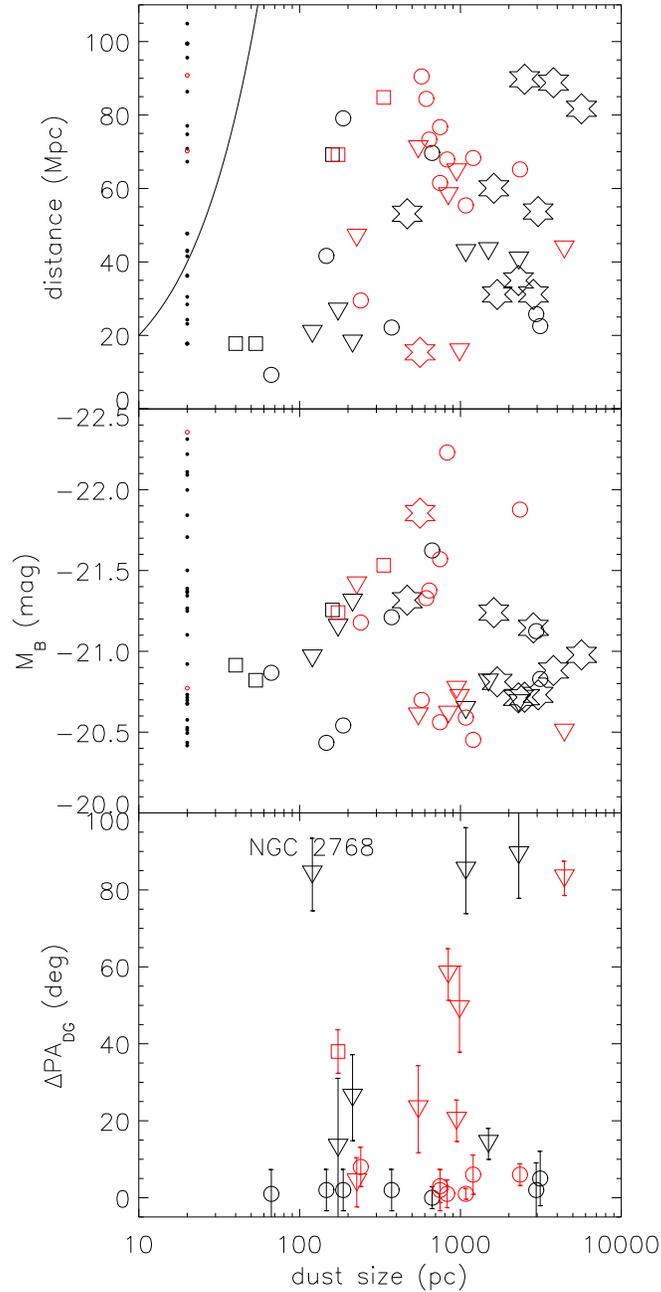}}
\caption{From top to bottom: galaxy
  distance, absolute blue magnitude and position angle difference
  $\Delta$PA$_{\rm DG}$ between host galaxy and central dust as a
  function of linear dust extent.  UGC~FR-I galaxies are represented
  by red symbols and UGC non-FR-I galaxies by black symbols. Circles
  indicate dust ellipses, triangles indicate dust lanes, squares
  indicate dust structures which have a morphology in between ellipse
  and lane and stars indicate irregular dust (see Section~\ref{s:dustproperties}). 
  The small dots indicate galaxies
  without dust detection plotted with a fiducial linear dust extent of
  $20\pc$. The typical relative error on dust size is $10\%$. The curve in the top panel indicates
  the minimum linear dust extent which can be resolved as a function
  of distance. In the bottom panel NGC 2768 is indicated because it has irregular dust with an extent of
  $\sim 1.7$ kpc roughly parallel to its small scale dust lane. See
  Section~\ref{s:dustcomparison} for further discussion.
}
\label{f:dustcomparison}
\end{centering}
\end{figure*}

We first check whether the observed dust properties in our galaxy samples depend on distance.
Figure~\ref{f:dustcomparison} compares the radio galaxies (the
UGC~FR-I sample) and non-radio galaxies (the UGC~non-FR-I sample). The
top diagram shows that galaxies with and without dust detections are
observed throughout the sampled volume. A similar range in dust sizes
is detected at every distance except for small ($<100\pc$) dust
distributions which HST can resolve well only at small distances.  We
see no trend in morphology classification with distance. We conclude
that the detection of dust, the classification and its size do not
correlate with galaxy distance for $v < 7000\kms$.

We detect dust in $48\% \pm 10\%$ of the UGC~non-FR-I galaxies. This
detection rate is similar to the $40\% \pm 9\%$ and $43\% \pm 8\%$
detection rates from HST imaging for the samples of nearby early-type
galaxies as compiled by van Dokkum \& Franx (\cite{vDok95})---excluding
galaxies with extended radio structures in their sample---and by Tran
\etal (\cite{Tra01}), respectively.  Both studies report an increased dust
detection rate for galaxies with radio emission (compact and/or
extended), but those rates are still less than the $90\% \pm 7\%$
detection rate for our UGC~FR-I sample of more powerful radio
galaxies. At much lower radio luminosities ($L \sim 10^{19-21}$W/Hz at
3.6cm), the radio luminosity functions of nearby early-type galaxies with and
without HST dust detection seem to become very similar (Kraj\-novi\'c \&
Jaffe \cite{Kra02}). Hence, there is an increasing dust detection rate with
increasing radio power. Many of the dusty non-radio galaxies
have low-level active galactic nuclei, sometimes with radio cores
(e.g., Ho \etal \cite{Ho97}; Tran \etal \cite{Tra01}). Thus the difference in
detection rate is perhaps a trivial effect: it could reflect the
simple fact that all radio galaxies have active galactic nuclei. The
close connection between dust and nuclear activity then supports the
idea that the central dusty ISM forms---not surprisingly---the fuel
reservoir for the active nucleus. 

All three dust morphologies occur in both radio and non-radio
galaxies. Dust features with a lane-like morphology appear in dusty
radio and non-radio galaxies at similar frequencies. In contrast,
irregular dust distributions are more often seen in non-radio galaxies
while radio galaxies harbour more often dust ellipses (see
Table~\ref{t:dustmorph}).  Thus more powerful radio sources occur
preferentially in galaxies with regular (perhaps more
'settled'?) dust.
 
Lanes and ellipses span the same range in size. There is no clear size difference for these dust morphologies in radio
and non-radio galaxies. The compact and/or faint features which could
be either lanes or ellipses form the smallest structures at a given
distance. The irregular dust distributions tend to be the largest dust
features. Larger-scale dust distributions tend to have a more
irregular morphology. \looseness=-2

The middle panel of Figure~\ref{f:dustcomparison} reveals no
systematic difference between the host magnitude of each dust class
for either the UGC~FR-I or the UGC~non-FR-I sample. Perhaps galaxies
brighter than $M_{\rm B} \sim -21.5$ do not contain dust with sizes
below $\sim 200$pc, but this result is not statistically significant.

The bottom panel of Figure~\ref{f:dustcomparison} shows a correlation
between dust orientation and morphology. The regular ellipses align
within $\sim 10\deg$ with the galaxy major axis, but the filamentary
dust lanes do not. This result holds for both radio and non-radio
galaxies and is also seen in the sample of lower-luminosity early-type
galaxies of Tran \etal (\cite{Tra01}). A corollary of these results is that
the morphological classification based on {\sl appearance} cannot be
due (only) to viewing angle, but reflects (at least partly) {\sl
intrinsic} differences between dust lanes and ellipses.

The dust lanes in radio galaxies have $\Delta$PA$_{\rm DG}$ values 
over the complete range, i.e., $\Delta$PA$_{\rm DG} \sim
[0\deg,90\deg]$. In contrast, dust lanes in non-radio galaxies are
within $\sim 25\deg$ from either the minor- or major axis. It is not
clear if this absence of large misalignment angles is simply due to
the limited size of the sample. A similar absence was also seen by Tran
\etal (\cite{Tra01}) from WFPC2 imaging, but not by van Dokkum \& Franx (\cite{vDok95})
using WFPC1 imaging. At the same time (and regardless of
morphology) smaller sized ($ \la 1\kpc$) dust structures tend to
align within $30\deg$ of the galaxy major axis. 
Interestingly, only kpc-scale dust lanes, and the 120pc dust lane in NGC 2768, are
roughly aligned with the galaxy minor axis. NGC 2768 has large-scale irregular dust with an extent of $\sim 1.7$ kpc
roughly along the minor axis of the galaxy, i.e., parallel to its small dust
lane. This irregular dust is associated with an extended disk of
emission-line gas which rotates around the major axis of the galaxy,
i.e., orthogonally to the main motion of the stars (McDermid \etal \cite{McD04};
Sarzi \etal \cite{Sar05}).
The
similarity between radio and non-radio galaxies in their relative
orientations of dust lanes and ellipses, suggests, if taken at face
value, that their orientations are not influenced by the presence of a
radio jet.


\section{Relative Position Angles of Dust, 
         Jet and Host in Radio Galaxies}
\label{s:dustjet}

We now focus on the position angle of dust lanes and ellipses relative
to (i) the radio jet axis and (ii) the galaxy major axis for the FR sample (see Table~\ref{t:fr1}).

\begin{figure*}[t]
\begin{centering}
\resizebox{1.0\hsize}{!}{\includegraphics{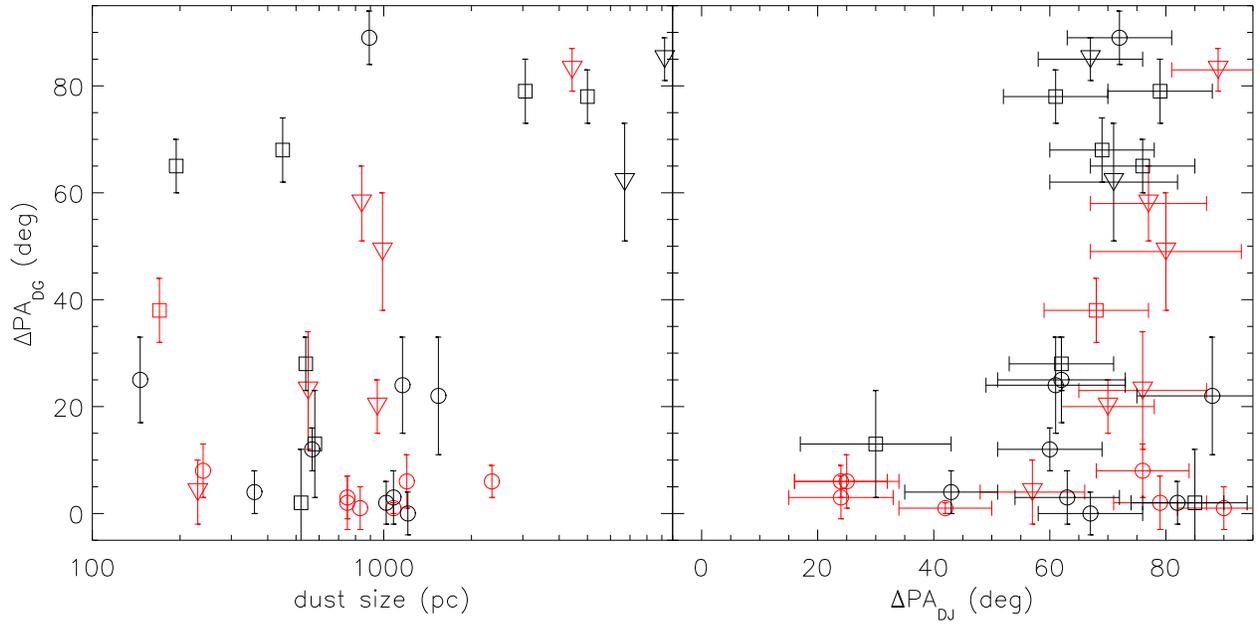}}
\caption{{\bf Left:} PA difference $\Delta$PA$_{\rm DG}$
  between dust axis and host galaxy major axis as a function of dust
  size for the FR sample. Galaxies with a dust ellipse, a dust lane
  or an 'intermediate' morphology, i.e., in between lane and ellipse, are
  denoted by circles, triangles and squares, respectively. The subset
  of UGC~FR-I galaxies have red symbols. The typical relative error on dust
  size is 10\%. The trends in the full FR sample are similar to
  those observed in the UGC~FR-I and UGC non-FR-I sample as shown in Fig.~\ref{f:dustcomparison}. {\bf Right:}
  $\Delta$PA$_{\rm DG}$ as a function of position angle difference
  between dust and radio jet $\Delta$PA$_{\rm DJ}$.}
\label{f:dpa}
\end{centering}
\end{figure*}

The left panel of Figure~\ref{f:dpa} is similar to the bottom panel of
Figure~\ref{f:dustcomparison}, but now for the FR sample. It
confirms the trends between the PA difference between
dust and galaxy major axis $\Delta$PA$_{\rm DG}$ and dust size of the
subset of UGC~FR-I galaxies in Figure~\ref{f:dustcomparison}. The segregation in the location of the data points as a
function of apparent dust morphology is now
less clear-cut. Dust ellipses still tend to align with the galaxy
major axis (i.e., $\Delta$PA$_{\rm DG} < 25\deg$), but the alignment
is not as tight as for the UGC~FR-I sample by itself. Furthermore, there is one notable exception at $\dpadg \sim 90\deg$ (3C 76.1, see Section~\ref{s:discussion} for further discussion). Again, most
lanes have large misalignments with the galaxy major axis (i.e.,
$\Delta$PA$_{\rm DG} > 20\deg$) and the largest lanes align
roughly with the galaxy minor axis.

The right panel of Figure~\ref{f:dpa} shows $\Delta$PA$_{\rm DG}$ as a function of the
PA difference between dust and jet axis $\Delta$PA$_{\rm DJ}$. We note
three special features. First, no radio jets are observed close to the
dust longest axis, i.e., $\Delta$PA$_{\rm DJ}<20\deg$. Second, the
data points are distributed roughly along a mirrored 'L' shape. Dust
structures which are roughly aligned with the galaxy major axis
($\Delta$PA$_{\rm DG} < 20\deg$) have a wide distribution in relative
angles with the radio jet ($\Delta$PA$_{\rm DJ}\sim [20\deg-90\deg]$).
In contrast, dust features which are misaligned from the galaxy major
axis ($\Delta$PA$_{\rm DG} > 20\deg$) have a narrow distribution in PA
differences with the radio jets ($\Delta$PA$_{\rm DJ}\sim
[60\deg-90\deg]$). These dust structures, while
misaligned with the galaxy, are in a very rough sense perpendicular to
the radio jets. Third, dust ellipses and lanes have a different
distribution of $\Delta$PA$_{\rm DJ}$. Ellipses have a wide
distribution, $\Delta$PA$_{\rm DJ} \sim 20\deg-90\deg$. In contrast,
lanes have a narrow distribution: all lanes except one have
$\Delta$PA$_{\rm DJ} \ga 65\deg$. The $\Delta$PA$_{\rm DJ}$ of the dust
lane galaxy 3C 353, for which no $\Delta$PA$_{\rm DG}$ is available
(Table~\ref{t:fr1}), is consistent with this.\looseness=-2

The FR sample contains 14 galaxies with dust features classified
as 'ambiguous between ellipse and lane' (cf.\ Table~\ref{t:fr1}).
$\Delta$PA$_{\rm DG}$ and $\Delta$PA$_{\rm DJ}$ can be measured for
eight of them, and are plotted in Figure~\ref{f:dpa}. Their distribution
among the lanes and ellipses supports the idea that in some cases
these dust features are ellipses and in some cases they are lanes.

What (selection) effect(s) could create the mirrored L shape in the
right panel of Figure~\ref{f:dpa}? It is not unreasonable to assume
that jets ploughing through a dusty medium would quickly destroy the
dust. This could explain the absence of systems with $\Delta$PA$_{\rm
DJ} < 20\deg$. Furthermore, suppose that jets are roughly
perpendicular to the dust plane {\sl intrinsically}. If the dust
structures are intrinsically circular and viewed close to
edge-on, then we will observe $\epsilon \sim 1$ in combination with $\Delta$PA$_{\rm
DJ} \sim 90\deg$. If such a circular disk is viewed close to face-on,
we will observe $\epsilon \sim 0$ in combination with a wide range of
relative position angles, i.e., $\Delta$PA$_{\rm DJ} \sim
[0\deg,90\deg]$. The thin filamentary dust lanes might be interpreted
as systems viewed relatively close to edge-on. Their high values of
$\Delta$PA$_{\rm DJ} \ga 60\deg$ are then qualitatively consistent
with radio jets being {\sl intrinsically} approximately perpendicular
to the dust plane.\looseness=-2

\section{Are the dust ellipses intrinsically circular?}
\label{s:ellipsecircular}

Before analysing the intrinsic orientations of dust ellipses and radio jets,
 we test if the observed distribution of $\epsilon$ is consistent with
circular or elliptical, thin or thick disks, assuming random viewing angles. This
assumption is invalid if (i) the disk detection rate or classification
depends on disk ellipticity, or (ii) the radio selection criteria
result in a non-random distribution of jet viewing angles {\sl and}
there is a relation between disk and jet orientation. Scenario (i) is
unlikely for the following reasons. We detect dust ellipses within
almost the full range of possible ellipticities (i.e., 0.01 to
0.86). There is no correlation between observed ellipticity and
distance. There is a dependence of classification on distance for dust structures classified as 'intermediate' (i.e.,
between ellipse and lane). At distances $D<150\Mpc$, $\sim 14\%$ of dusty galaxies in the radio and non-radio galaxy samples are classified as intermediate. These do not have a clear bias for $\epsilon$ greater or smaller
than 0.5. However, nine of the eleven galaxies
at $D>150\Mpc$ have an ambiguous dust morphology classification of
which six have $\epsilon \ga 0.5$ (see Table~\ref{t:fr1}). At these
large distances it becomes impossible with the current imaging to
distinguish unambiguously between high-ellipticity ellipses and lanes
in general. Thus we limit the analysis of disk shapes to galaxies at
$D<150\Mpc$.  Scenario (ii) does not seem to be relevant for the
UGC~FR-I sample as it is selected on total 1400 MHz flux which is
dominated by the extended and unbeamed radio-lobe emission. The radio
cores from VLA 1490 MHz measurements (FWHM $\sim 1.5''-3.75''$)
constitute always less than $22\%$ of the total 1400 MHz flux and
typically $\sim 6\%$ (Xu \etal \cite{Xu00}). Radio galaxies in the 3CR sample
and B2 sample are selected using total 178 MHz and 408 MHz fluxes,
respectively (Bennett \cite{Ben62}; Colla et al., \cite{Col70}). At these lower
frequencies the radio emission is expected to be even more lobe-dominated.

\begin{figure*}[t]
\begin{centering}
\resizebox{1.0\hsize}{!}{\includegraphics{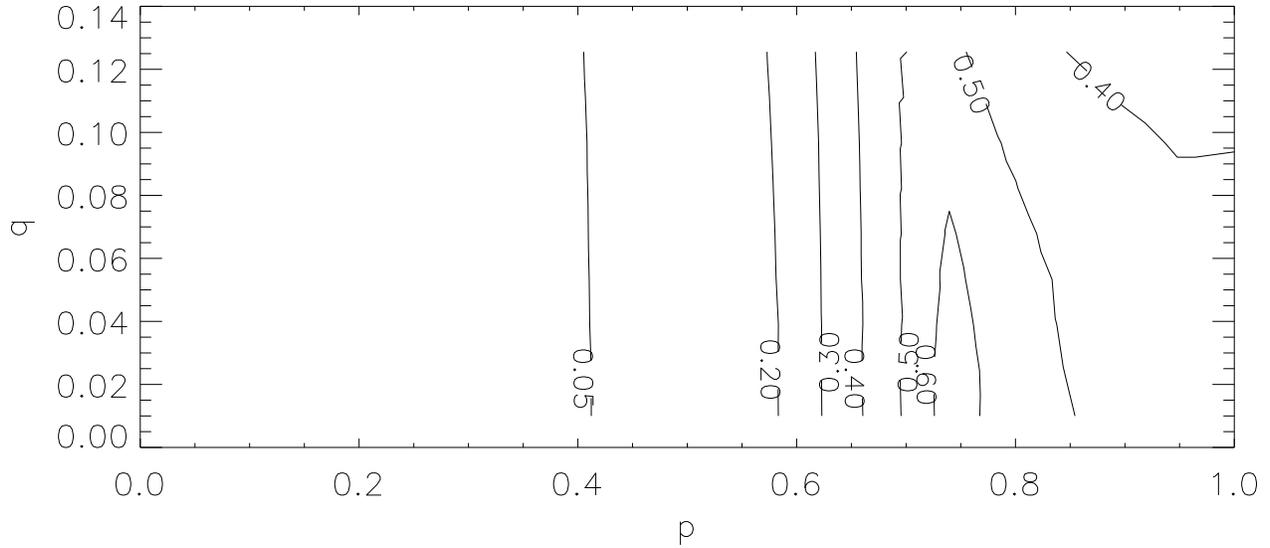}}
\caption{
Contours of constant probability using a Kolmogorov-Smirnov (KS) test for the triaxial properties of
the dust ellipses. The hypothesis is that the ellipticities $\epsilon$ of the dust ellipses, observed in 18
F-I and FR-I/II galaxies at $D<150\Mpc$, are caused by randomly-oriented triaxial bodies with an
intermediate-to-long axis ratio $p$ and short-to-long axis ratio $q$. Each contour is labelled with its KS
probability. The distribution of $\epsilon$ in the radio galaxies requires relatively thin ($q<0.14$) disks
and is most consistent with close to circular disks ($p >\sim 0.7$). A similar result is obtained for the
combined sample of 25 non-radio and radio galaxies with dust ellipses at $D<150\Mpc$. 
See Section~\ref{s:ellipsecircular} for details.
}
\label{f:kseps}
\end{centering}
\end{figure*}

We model dust ellipses as randomly-oriented, intrinsically triaxial bodies with an
intermediate-to-long axis ratio $p$ and short-to-long axis ratio
$q\leq p \leq 1$. Thus an infinitely thin circular disk has $p=1$ and
$q=0$.  We compare the observed distribution of $\epsilon$ to the
expected distribution for given $p, q$ using the Kolmogorov-Smirnov
(KS) test. Figure~\ref{f:kseps} shows the contour levels of equal KS probability in the plane of the axis ratios $p$ and $q$, for the sample of 18 radio galaxies with dust ellipses at $D<150\Mpc$. The maximum $q$ allowed by the data is $q=0.14$, given that the maximum observed ellipticity is $\epsilon=0.86$. The contours are almost independent of $q$, suggesting that the data set is too small to put any further constraint on $q$. The KS probability is a function of $p$ and maximizes for close to oblate ($p \sim 0.74$) disks. Similar results are obtained for the combined sample of 25 non-radio and radio galaxies at $D<150\Mpc$ with dust ellipses. In the analysis of the three dimensional configuration of jets and dust disks, we will consider three representative models: a thin circular disk ($p=1$, $q=0$), a thick circular disk ($p=1$, $q=0.13$) and a thin elliptic disk ($p=0.75$, $q=0$).

\section{Intrinsic orientation of dust ellipses and radio jets}
\label{s:ellipsejetintrinsic}

\begin{figure*}[t]
\begin{centering}
\resizebox{1.0\hsize}{!}{\includegraphics{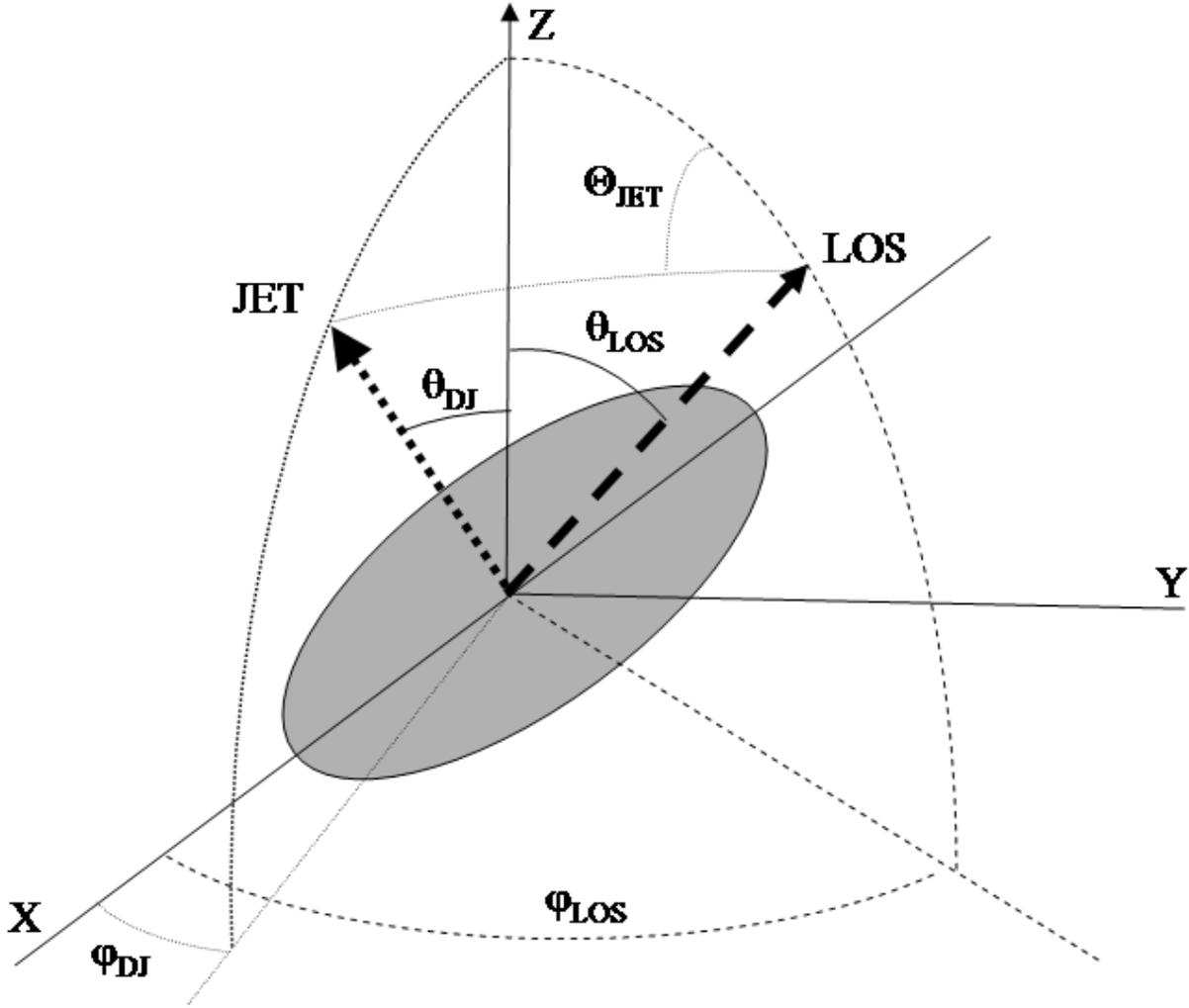}}
\caption{The coordinate system used to describe the dust -- jet system. The dust structure is assumed to be centred on the galaxy nucleus and to have a triaxial shape. The long, intermediate and short axis of the dust  distribution lie along the X, Y and Z-axis respectively. The jet axis is indicated by the dotted arrow and makes an angle $\theta_{\rm DJ}$ with the Z-axis and an azimuthal angle $\phi_{\rm DJ}$ with the X-axis. The line-of-sight direction, indicated by the dashed arrow, makes and angle $\theta_{\rm los}$ with the Z- axis and an azimuthal angle $\phi_{\rm los}$ with the X-axis. The spherical angle labelled $\Theta_{\rm jet}$ indicates the angle between the jet and the short axis projected in the plane of the sky. See Section~\ref{s:modelparameters} for formulae relating these quantities.}
\label{f:angledef}
\end{centering}
\end{figure*}

We now attempt to constrain the {\sl intrinsic} orientation of the
jet-axis relative to the dust systems classified as ellipses in the
FR sample. 
Figure~\ref{f:angledef} illustrates the model parameters and
coordinate system which we will use to describe the dust -- jet
system. The long, intermediate and short
axes of the triaxial dust structure are chosen to lie along the X, Y and Z-axes, respectively. The
radio-jet axis makes an angle $\theta_{\rm DJ}$ with the Z-axis and an
azimuthal angle $\phi_{\rm DJ}$ with the X-axis. The line of sight
makes and angle $\theta_{\rm los}$ with the Z-axis (i.e., the disk
inclination angle) and an azimuthal angle $\phi_{\rm los}$ with the
X-axis. Thus, the disk-jet model has in general six parameters.

\subsection{The relation between model parameters and observables}
\label{s:modelparameters}

The observations provide only two parameters of the dust-jet system as
projected on the plane of the sky: the position angle difference
between jet and disk $\Delta$PA$_{\rm DJ}$ and the disk ellipticity
$\epsilon$. The ellipticity can be expressed in the model parameters
as follows (Contopoulos \cite{Con56}; Franx \cite{Fra88}):\looseness=-2
\begin{equation}
(1-\epsilon)^2= {a-\sqrt{b} \over a+\sqrt{b}},
\label{e:epsilon}
\end{equation}
with
\begin{eqnarray}
\begin{array}{ll}
 a =&(1-q^2)\cos^2\theta_{\rm los}+(1-p^2)\sin^2\theta_{\rm los}
      \sin^2\phi_{\rm los} +p^2+q^2, \\
 b =&[(1-q^2)\cos^2\theta_{\rm los}-(1-p^2)\sin^2\theta_{\rm los}
      \sin^2\phi_{\rm los} -p^2+q^2]^2 \\ 
    & \qquad\qquad 
      +4(1-p^2)(1-q^2)\sin^2\theta_{\rm los}\cos^2\theta_{\rm los}
                 \sin^2\phi_{\rm los}.  \\
\end{array}
\end{eqnarray}
The position angle $\Theta_{\rm min}$ of the minor axis of the dust
structure relative to the projection of the Z-axis (i.e., short axis)
can be written as:
\begin{equation}
\tan 2 \Theta_{\rm min} = 
    {2T\sin\phi_{\rm los}\cos\phi_{\rm los}\cos\theta_{\rm los} \over
         \sin^2\theta_{\rm los}-T(\cos^2\phi_{\rm los}-\sin^2\phi_{\rm los}
           \cos^2\theta_{\rm los})}, 
\label{e:thetaminor}
\end{equation}
with triaxiality parameter $T=(1-p^2)/(1-q^2)$. The value of
$\Theta_{\rm min}$ indicates the position angle of the minor (rather
than the major) axis if:
\begin{equation}
{\rm sign}(\tan\Theta_{\rm min}) = 
{\rm sign}(\sin\phi_{\rm los}\cos\phi_{\rm los}\cos\theta_{\rm los}).
\label{e:thetaminorsign}
\end{equation}
The position angle $\Theta_{\rm jet}$ of the jet-axis relative to the
position angle of the Z-axis can be expressed as (e.g., de Zeeuw \&
Franx \cite{dZee89}):
\begin{equation}
\tan \Theta_{\rm jet} = {\sin\Delta\phi\tan\theta_{\rm DJ} \over
                            \sin\theta_{\rm los} - 
                    \cos\Delta\phi\tan\theta_{\rm DJ} \cos\theta_{\rm los}}, 
\label{e:thetajet}
\end{equation}
with $\Delta\phi=\phi_{\rm los} - \phi_{\rm DJ}$. The position angle
difference $0\deg \leq \Delta{\rm PA}_{\rm DJ} \leq 90\deg$ can then
be expressed as:
\begin{equation}
\cos\Delta{\rm PA}_{\rm DJ} = \sin(90\deg-\Delta{\rm PA}_{\rm DJ}) 
              = \bigl|\sin(\Theta_{\rm jet} - \Theta_{\rm min})\bigr|.
\label{e:dpadj}
\end{equation}
Eq.~\ref{e:thetaminor} shows that $\Theta_{\rm min}=0\deg$ for a
circular disk ($p=1$ and $q=0$), so that $|\sin\Theta_{\rm jet}|$
corresponds to $\sin(90\deg-\Delta$PA$_{\rm DJ})$.  The expressions for $\epsilon$ and
$\Delta$PA$_{\rm DJ}$ then simplify to:
\begin{equation}
\epsilon=1-\bigl| \cos\theta_{\rm los} \bigr|,
\label{e:epsiloncircle}
\end{equation}
and
\begin{equation}
\tan(90\deg-\Delta{\rm PA}_{\rm DJ})= 
   \Bigl|\frac{\sin\Delta\phi \tan\theta_{\rm DJ}}
   {\sin\theta_{\rm los} 
              - \cos\Delta\phi\tan\theta_{\rm DJ} \cos\theta_{\rm los}}\Bigr|.
\label{e:dpadjcircle}
\end{equation}
Thus $\Delta{\rm PA}_{\rm DJ}=90\deg$ for any line of sight if
$\theta_{\rm DJ}=0\deg$.

\subsection{Lower limits on $\theta_{\rm DJ}$}
\label{s:lowerlimitsthetadjcircle}

The angle between the line of sight and the radio jet is
generally unknown. For an oblate dust structure (i.e., $p=1$ and $\Theta_{\rm min}=0\deg$), the observed position angle difference with
the disk merely confines the jet orientation to lie anywhere on the
two 'jet-circles' which are the two great circles defined by the
line-of-sight vector and the angle $\Theta_{\rm jet}$ (cf.\
Fig.~\ref{f:angledef} and Eq.~\ref{e:dpadj}; there are two jet-circles because
$\Delta$PA$_{\rm DJ}$ is an absolute value). The values of
$\theta_{\rm DJ}$ and the difference in azimuthal angle $\Delta\phi$
fix the jet location in the jet-circles. We will refer to $\theta_{\rm
DJ}$ as the 'misalignment angle'. To observe simultaneously a given
position angle difference $\Delta$PA$_{\rm DJ} < 90\deg$ and disk
ellipticity $\epsilon$ requires a minimum misalignment angle
$\theta_{\rm DJ}^{\rm min}$. Geometrically, this angle defines the
latitude on a unit sphere which is tangent to the jet-circles. The jet
axis intersects one of the two tangent points. The minimum
misalignment angle can be expressed as (e.g., Schmitt \etal \cite{Sch02})
\begin{equation} 
\sin\theta_{\rm DJ}^{\rm min} = 
                    \cos\Delta{\rm PA}_{\rm DJ}\sin\theta_{\rm los}.
\label{e:thetamin}
\end{equation}
Since most radio galaxies have $\Delta$PA$_{\rm DJ} \neq 90\deg$ (Table~\ref{t:fr1}), it follows that jets typically have
$\theta_{\rm DJ}^{\rm min} > 0\deg$, irrespective of our line of sight
towards the system. The $\theta^{\rm min}_{\rm DJ}$ for circular/oblate dust disks are listed in
Table~\ref{t:fr1}. The maximum $\theta^{\rm min}_{\rm DJ}$ required by
the FR sample is $54\deg$ for 3C 449 and four of the 16 disks for
which $\Delta$PA$_{\rm DJ}$ is available have $\theta^{\rm min}_{\rm
DJ} > 40\deg$. If the disks are circular or oblate, significant
misalignments occur between jet axis and disk normal.

\subsection{The distribution of $\theta_{\rm DJ}$: thin disks}
\label{s:mlellipses}

\begin{figure*}[t]
\begin{centering}
\resizebox{1.0\hsize}{!}{\includegraphics{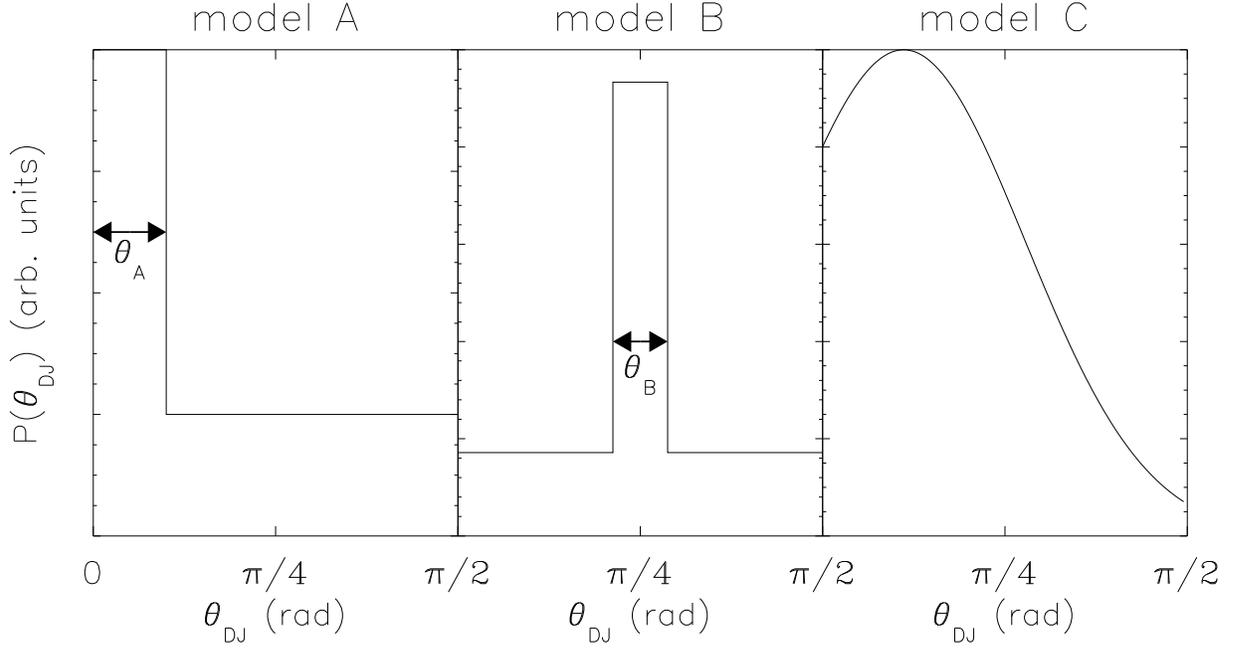}}
\caption{The three probability distributions of the jet / dust disk misalignment angle
$\theta_{\rm DJ}$ assumed to underlie the observations. {\bf Left:} model A, the single-step function as
described by Equation~\ref{e:singlestep}. The width and height of the level at low $\theta_{\rm DJ}$ depend
on the free parameter $\theta_A$. The integrated probability underneath this level is $1-2\theta_A/\pi$. This
model is used to test if the observations imply a peak in the $\theta_{\rm DJ}$ distribution at low or high
$\theta_{\rm DJ}$. {\bf Middle:} model B, the double-step function as described by
Equation~\ref{e:doublestep}. The width and height of the central level are set by $\theta_B$ and is centred
on $\theta_{\rm DJ}=\pi/4$. The integrated probability under the central level is $1-2\theta_B/\pi$. Model B
tests if the observations are consistent with a central peak or dip in the distribution of $\theta_{\rm DJ}$.
{\bf Right:} model C, a truncated Gaussian normalized to 1 (see Eq.~\ref{e:gaussian}). Both the mean
$0\leq \mu \leq \pi/2$ and dispersion $0\leq \sigma \leq \pi$ are free parameters. This model tests for the
presence of a peak anywhere in the range $0 \leq \theta_{\rm DJ} \leq \pi/2$. See Section~\ref{s:mlellipses}
for details.}
\label{f:modelthetadj}
\end{centering}
\end{figure*}

We have lower limits on $\theta_{\rm DJ}$ for individual circular
disk-jet systems.  The next step is to constrain the full distribution
of $\theta_{\rm DJ}$. This constitutes an inversion problem: we have
to recover the distribution functions of misalignment angles
$\theta_{\rm DJ}$, azimuthal angles $\phi_{\rm DJ}$, and line-of-sight
directions $(\theta_{\rm los},\phi_{\rm los})$ from the sample of
observational pairs ($\Delta$PA$_{\rm DJ}$,$\epsilon$).  We limit the analysis to the 18 dust ellipses at $D<150\Mpc$ which are consistent with a
spherically uniform random distribution of the line of sights (cf.\
Section~\ref{s:ellipsecircular}). We assume that $\phi_{\rm DJ}$ has a uniform random distribution. The relative intrinsic orientation between jet and dust is
characterized then by the misalignment angle distribution function $P_{\rm
DJ}(\theta_{\rm DJ})$. We interpret the distribution function as
a probability density function. Given the small number of
observations, we explore three parameterizations for $P_{\rm DJ}$ instead of
attempting a parameter-free recovery. They are illustrated in Figure~\ref{f:modelthetadj}. First, the observations
are interpreted as being drawn from a step-function probability density distribution
$P_{\rm DJ}(\theta_{\rm DJ})$ (angles in radians):
\begin{eqnarray}
\label{e:singlestep}
\begin{array}{ll}
 P_{\rm DJ}(\theta_{\rm DJ})d\theta_{\rm DJ} =\frac{2}{\pi}\frac{\frac{\pi}{2} - \theta_A}{\theta_A}d\theta_{\rm DJ} , \qquad 0 \leq \theta_{\rm DJ} \leq \theta_A, \\
 P_{\rm DJ}(\theta_{\rm DJ})d\theta_{\rm DJ} =\frac{2}{\pi}\frac{\theta_A}{\frac{\pi}{2}-\theta_A}d\theta_{\rm DJ} , \qquad otherwise. 
\end{array}
\end{eqnarray}
The angle $\theta_A$ is a free parameter. This step function, which we call model A, tests the hypothesis that the distribution of $\theta_{\rm DJ}$ is peaked either at 
small or large misalignment angles or better agrees with a uniform distribution (i.e., $\theta_A=\pi/4$). Quantitatively, a fraction $1-2\theta_A/\pi$ of the jets have misalignment angles $\theta_{\rm DJ} < \theta_A$, while the average $\theta_{\rm DJ}=\theta_A$.  
Thus a $\theta_A=\pi/4$ corresponds to a uniform distribution in $\theta_{\rm DJ}$. Second, to test the hypothesis that the distribution of $\theta_{\rm DJ}$ does peak at intermediate misalignment angles (i.e., $\theta_{\rm DJ} = \pi/4$) we analyse 
the data with the following two-step function, which we call model B:
\begin{eqnarray}
\label{e:doublestep}
\begin{array}{ll}
 P_{\rm DJ}(\theta_{\rm DJ})d\theta_{\rm DJ} =\frac{2}{\pi}\frac{\frac{\pi}{2} - \theta_B}{\theta_B}d\theta_{\rm DJ}, \qquad \frac{\pi}{4}-\frac{\theta_B}{2} \leq \theta_{\rm DJ} \leq \frac{\pi}{4}+\frac{\theta_B}{2}, \\
 P_{\rm DJ}(\theta_{\rm DJ})d\theta_{\rm DJ} = \frac{2}{\pi}\frac{\theta_B}{\frac{\pi}{2}-\theta_B}d\theta_{\rm DJ}, \qquad otherwise. 
\end{array}
\end{eqnarray}
In this case a fraction of $1-2\theta_B/\pi$ jets have $\pi/4-\theta_B/2 < \theta_{\rm DJ} < \pi/4+\theta_B/2$ and the average $\theta_{\rm DJ}=\pi/4$, regardless of $\theta_B$. 
Thus $\theta_B=\pi/4$ corresponds to a uniform distribution in $\theta_{\rm DJ}$. Third, we explore a Gaussian distribution for $\theta_{\rm DJ}$:
\begin{equation}
\label{e:gaussian}
P_{\rm DJ}(\theta_{\rm DJ})d\theta_{\rm DJ} = C \exp(-\frac{1}{2}\frac{(\theta_{\rm DJ} - \mu)^2}{\sigma^2})d\theta_{\rm DJ}, \qquad  0 \leq \theta_{\rm DJ} \leq \frac{\pi}{2}
\end{equation}
The Gaussian is truncated to the physically allowed region $0 \leq \theta_{\rm DJ} \leq \pi/2$ and normalized to 1 using constant $C$. Both mean $\mu$ and dispersion $\sigma$ (of the full Gaussian) are left free to vary in the ranges $0\leq \mu \leq \pi/2$ and $0 \leq \sigma \leq \pi$. This model, which we call model C, explores the presence of a peak not just at $\theta_{\rm DJ} = \pi/4$ but anywhere and with a free width. 


First, we estimate the $\theta_A$
that best fits the observations using a Maximum Likelihood
analysis. The goal is to maximize the likelihood $L$ of observing our data set of $\Delta{\rm PA}_{\rm DJ}$ and $\epsilon$:
\begin{equation} 
L=\prod\limits_{i=1}^{n_{\rm obs}} P(\Delta{\rm PA}_{{\rm DJ},i} \, ,
                   \epsilon_i|\theta_A)d(\Delta{\rm PA}_{\rm DJ}) d\epsilon, 
\end{equation}
where the product is over the $n_{\rm obs}$ observations.  We
approximate $P(\Delta{\rm PA}_{\rm DJ}, \epsilon|\theta_A)$ using a
Monte-Carlo simulation. We draw a large number of realizations of
random line of sights $(\theta_{{\rm los}},\phi_{{\rm los}})$, random
azimuthal jet angles $\phi_{{\rm jet}}$ and misalignment angles
$\theta_{\rm DJ}$ from model A. Using
Eqs~(\ref{e:epsiloncircle}) and (\ref{e:dpadjcircle}), these
realizations are expressed in $\Delta{\rm PA}_{\rm DJ}$ and
$\epsilon$. A two-dimensional histogram of the realizations with
$n_{\rm bin}^2$ equal-sized bins approximates $P(\Delta{\rm PA}_{{\rm
DJ},i},\epsilon_i|\theta_A)$ and hence the likelihood. The likelihood is
maximized using an amoeba routine (Press \etal \cite{Pre92}) with $\theta_A$ as
the optimising parameter maximizing $L$. We choose the numerical
simulation approach for two reasons. First, $P(\Delta{\rm PA}_{\rm
DJ}, \epsilon)$ is not available in analytical form in
general. Second, the Monte-Carlo simulation allows us to incorporate
easily observational selection effects. For
example, the three dust ellipses with $\epsilon < 0.1$ are too
roundish to determine $\Delta$PA$_{\rm DJ}$ (see
Section~\ref{s:dustproperties}). We account for this by selecting a
subset of $n_{\rm mc}$ pairs of ($\Delta$PA$_{\rm DJ}$,$\epsilon$)
from the Monte-Carlo realizations which have $\epsilon>0.1$. We
performed simulations using $n_{\rm mc}=10^6$ and $n_{\rm
bin}=10$. The numerical error arising from this choice is negligible compared to the uncertainties in $\theta_A$ caused by the limited size and accuracy of the data set. Maximum Likelihood estimators are not guaranteed to be
unbiased (e.g., Cowan \cite{Cow98}). We verified with simulations that
the bias of the Maximum Likelihood estimator used here does not affect our results.

The Maximum Likelihood analysis infers $\theta_A^{\rm obs}=43\deg$ for
the $n_{\rm obs}=15$ observations with both $\Delta$PA$_{\rm DJ}$ and
$\epsilon>0.1$. We verify that the data are plausibly drawn from the
best-fitting model by comparing the value of the Maximum Likelihood $L^{\rm
max}_{\rm obs}$ of the observations to $L^{\rm max}_{\rm sim}$ of
simulated data sets which were created using the best-fitting
$\theta_A^{\rm obs}$. These mock samples contain the same number of
galaxies as the observations. 
We find $L^{\rm max}_{\rm obs} > L^{\rm
max}_{\rm sim}$ in 60\% of the mock samples and hence the hypothesis is statistically well
accepted. \looseness=-2

We estimated the confidence levels on $\theta_A^{\rm obs}$ in a similar way. Figure~\ref{f:confidence} (top-left) shows the cumulative distribution function of $\theta_A^{\rm sim}$ inferred from 100 mock samples which are created in three ways:
\begin{enumerate}
\itemsep 0pt
\item{In a Monte-Carlo fashion using the best-fitting $\theta_A^{\rm obs}$ in
      model A from the observations and the requirement that $\epsilon > 0.1$.}
\item{'Dithering' the observations within the measurement errors.}
\item{By a bootstrap method, i.e., creating mock data sets of 
      ${\Delta{\rm PA}_{\rm DJ},\epsilon}$ by drawing $n_{\rm obs}$
      independent data pairs $(\Delta{\rm PA}_{\rm DJ},\epsilon)$ from
      the $n_{\rm obs}$ observations.}
\end{enumerate}
The
first method samples the uncertainty of the best fit due to the finite
number of observations under the assumption that the underlying model
is correct. The second method samples the uncertainty due to the
random observational measurement errors. The third method samples a
similar uncertainty as method 1 but with the observed data set itself
used as an estimator of the underlying probability distribution.

All three methods infer median values for $\theta_A^{\rm sim} \sim 40\deg-45\deg$ in good agreement with the observations. The 'dithering' method yields $1\sigma, 2\sigma$ confidence
intervals which are significantly smaller than those of the other
two methods. This shows that the statistical
spread in $\theta_A$ due to the finite
sample size is the dominant source of uncertainty for this model.  The confidence
intervals from the Monte-Carlo and bootstrapping method are more similar, although  the Monte-Carlo method shows the largest 'probability wings' at small and large misalignment angles. Therefore we take the result from the Monte-Carlo method as 
a conservative estimate of the confidence levels.

We conclude that the observed jet-disk relative position angles and
disk inclinations are consistent with a spherically random distribution of misalignment angles. The $1\sigma$ upper and lower
confidence levels indicate that the average $\theta_{\rm DJ}$ lies in the range $38\deg-56\deg$.

\subsection{The distribution of $\theta_{\rm DJ}$: thick disks}
\label{s:mlellipsesthick}

In Section~\ref{s:ellipsecircular} we showed that the dust ellipses
are not only consistent with thin circular disks (i.e., axis ratios
$p=1$ and $q=0$), but also with oblate (i.e., thickened) disks, with
axis ratios $p=1$ and $q=0.13$. For such structures,
Eq.~(\ref{e:epsilon}) simplifies to
\begin{equation}
\epsilon=1-\sqrt{\cos^2 \theta_{\rm los}+q^2 \sin^2 \theta_{\rm los}}
\label{e:epsilonoblate}
\end{equation}
Thus, in comparison to thin disks, the thick disks appear rounder at
any inclination $\theta_{\rm los}$. As a result, the inferred
inclinations for the disks in the radio galaxy sample increase under
the assumption of thick disks. Eq.~(\ref{e:thetamin}) then implies
that the minimum misalignment angle $\theta_{\rm DJ}^{\rm min}$
increases for each disk. We redid the Maximum Likelihood analysis
assuming dusty oblates with $q=0.13$. The analysis for thick disks
yields as expected a slightly larger, but very similar $\theta_A^{\rm obs}=43.5\deg$ compared to thin disks. Also the confidence levels do not change significantly (see Figure~\ref{f:confidence}). Thus the
inferred distribution of jet misalignment angles does not depend
critically on the assumed thickness of the disks.

\subsection{The distribution of $\theta_{\rm DJ}$: elliptic disks}
\label{s:mlellipseselliptic}

We showed in Section~\ref{s:ellipsecircular} that the dust ellipses
are also consistent with thin elliptic disks (i.e., $p=0.75$ and
$q=0$) observed at random viewing angles. In contrast to circular
disks, a $\Delta$PA$_{\rm DJ} < 90\deg$ can be observed for certain
viewing angles also for $\theta_{\rm DJ}=0\deg$. However, only a
fraction of the full range $0\deg \leq \Delta$PA$_{\rm DJ} \leq
90\deg$ can be observed for certain $\epsilon$.  Thus some dust
ellipses still require a minimum misalignment angle $\theta_{\rm
DJ}^{\rm min}>0\deg$ in the case of an elliptic disk. We determined
these angles numerically and they are listed in Table~\ref{t:fr1} next
to those for the case of a circular disk.  For three of the 16 dust
ellipses $\theta^{\rm min}_{\rm DJ}\geq 40\deg$ assuming
$p=0.75$. Thus, significant misalignments do occur also if the dust
ellipses are thin elliptic disks intrinsically.

We repeated the Maximum Likelihood analysis assuming that all dust
ellipses are elliptic disks with $p=0.75$ and $q=0$. The result is
plotted in the middle-left diagram in Figure~\ref{f:confidence}. As expected, the best-fitting
average misalignment angle $\theta_A^{\rm obs}=40\deg$ is smaller than for thin
circular disks. The change falls within the
$1\sigma$ confidence level for circular disks. Thus, the best-fit to
the distribution of jet misalignment angles does not depend critically
on the assumed ellipticity of the disks within the accepted range
$p=[0.75,1]$.

\subsection{The peak in the distribution of misalignment angles}
\label{s:compareab}

\begin{figure*}[t]
\begin{centering}
\resizebox{1.0\hsize}{!}{\includegraphics{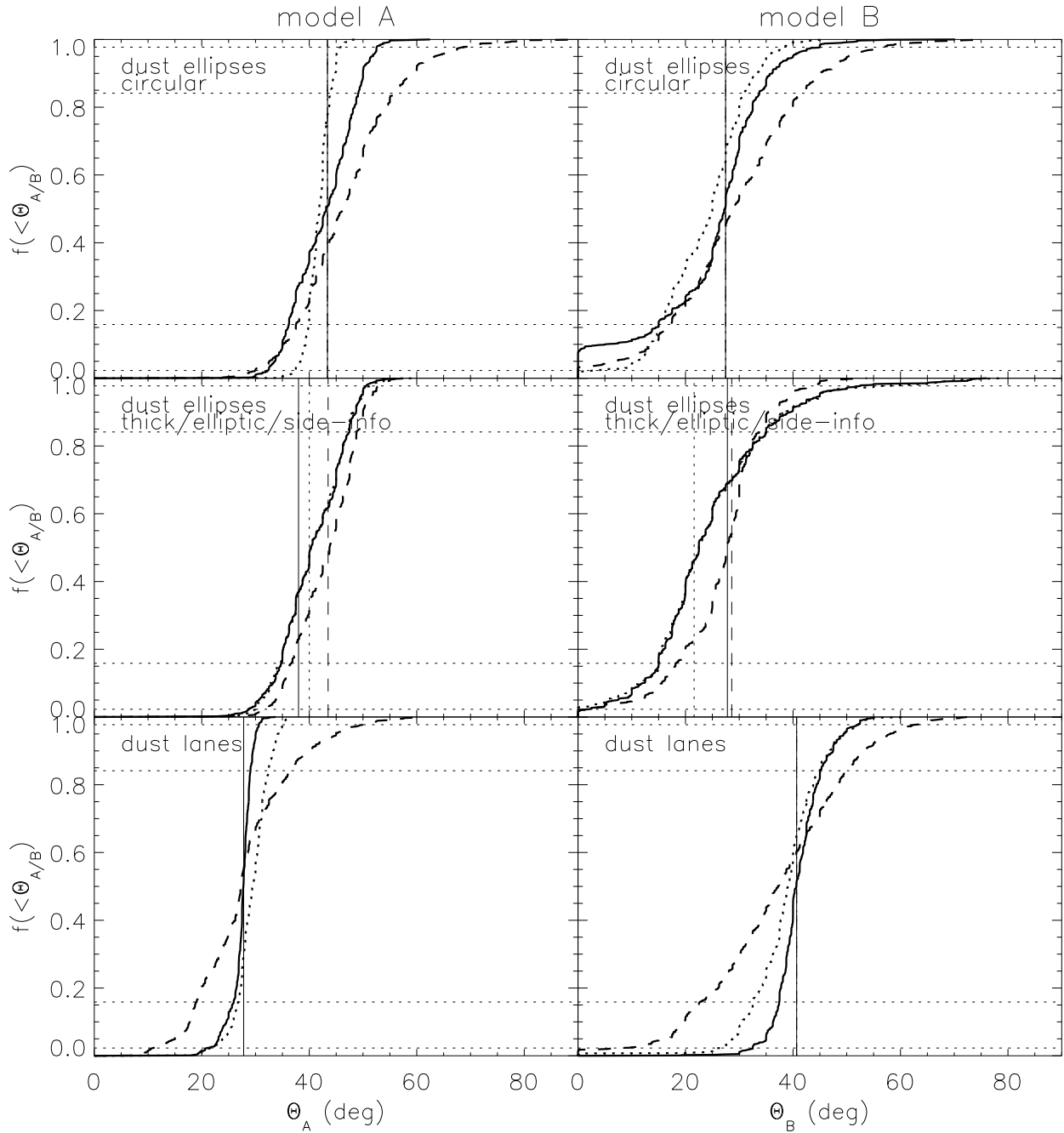}}
\caption{Confidence levels on the best-fitted parameter $\theta_A$ for model A (left
column)and $\theta_B$ for model B (right column) of the jet-dust misalignment angle, $\theta_{\rm DJ}$,
distribution. Plotted are cumulative distributions of the best-fitted free parameter, $\theta_A$ and
$\theta_B$, for various models of errors and dust morphology. The first four plots are models for the dust
ellipses, the bottom two for the dust lanes. {\bf Top}: thin circular disks are assumed ($p=1$ and $q=0$). To
estimate the error on $\theta_A$ the bootstrap (solid), Monte-Carlo (dashed) and dither (dotted) method are
used (see Section~\ref{s:ellipsejetintrinsic}). The two-sided $68.27\%$ and $95.45\%$ confidence levels
around the median are indicated by the horizontal dotted lines. The vertical lines indicate the best-fitted
$\theta_A$ and $\theta_B$ for the observations. {\bf Middle}: similar to the top model, but now for oblate
'thick' disks ($p=1$, $q=0.13$, dashed curve) and thin elliptic disks ($p=0.75$, $q=0$, dotted curve) and for
thin circular disks ($p=1$, $q=0$, solid curve). The latter model uses additional information on the near
side of jet and disk which is available for 10 of the 15 galaxies. All curves use the bootstrap method for
error estimation. {\bf Bottom}: similar to top diagrams, but now for eight lane galaxies. All models assume
edge-on disks. The three methods of error estimation are indicated by the same line style as in the top
diagrams. The main conclusion of the six diagrams is that dust lanes are most
consistent with smaller misalignment angles than dust ellipses. The difference in misalignment angle is more clearly inferred by model C as shown
in Figure~\ref{f:cdfthetadj}. See Sections~\ref{s:ellipsejetintrinsic} and~\ref{s:lanejetintrinsic} for a
detailed discussion.}
\label{f:confidence}
\end{centering}
\end{figure*}

The absence of peaks at small or large misalignment angles resulting from using model A cannot rule out a best-fit distribution of misalignment angles which peaks at an intermediate $\theta_{\rm DJ} \sim 45\deg$. Thus it is worthwhile to explore model B which can test for such peaks. For circular disks,
the analysis infers a best-fitting $\theta_B^{\rm obs}= 27\deg$, i.e., a broad peak corresponding to$\sim 70\%$ of the jets having misalignment angles $\theta_{\rm DJ}=[32\deg,58\deg]$. Figure~\ref{f:confidence} (top-right) shows the confidence levels using the three methods of parameter error estimation also used in Section~\ref{s:mlellipses}. Although a broad peak is preferred, all methods indicate that neither a single $\theta_{\rm DJ}=45\deg$ for all jets nor a distribution uniform in $\theta_{\rm DJ}$ (i.e., $\theta_B=45\deg$) can be ruled out at more than the $95\%$ confidence level.

Figure~\ref{f:confidence} (middle-right) shows that the assumption of thick disks yields similar preferred $\theta_B$, while assuming thin elliptic disks yields $\theta_B=22\deg$, i.e., a peak which is $5\deg$ smaller in width. However the upper $95\%$ confidence level on $\theta_B$ increases, while small $\theta_B \la 10\deg$ are ruled out at larger confidence compared to thin circular disks.

\begin{figure*}[t]
\begin{centering}
\resizebox{1.0\hsize}{!}{\includegraphics{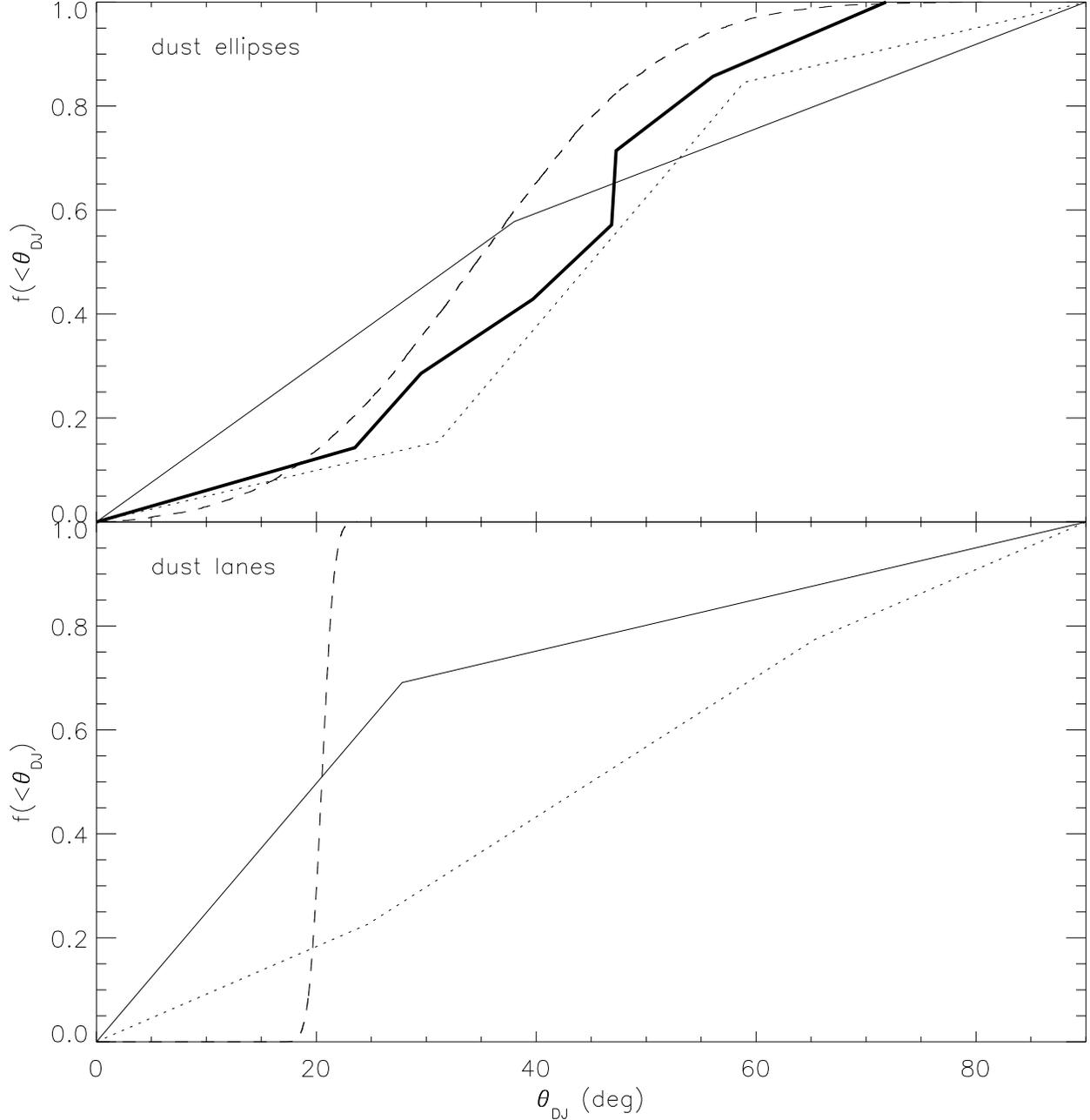}}
\caption{Cumulative distributions of the jet-dust misalignment angle $\theta_{\rm DJ}$ for
  the radio galaxies with dust ellipses (top) and dust lanes (bottom). {\bf Top:} the three thin curves
  indicate the cumulative distributions obtained from the Maximum Likelihood analyses of the observed
  distribution of dust ellipticities $\epsilon$ and dust-jet position angles differences $\Delta$PA
  $_{\rm DJ}$ from 15 FR-I and FR-I/II radio galaxies (see Section~\ref{s:ellipsejetintrinsic}). The solid
  curve corresponds to model A (i.e., single-step function, cf.\ Equation~\ref{e:singlestep}), the dotted
  curve corresponds to model B (i.e., two-step function, cf.\ Equation~\ref{e:doublestep}) and the dashed
  curve for model C (i.e., the truncated Gaussian distribution, cf.\ Equation~\ref{e:gaussian}). The models
  assume thin circular dust disks and take into account the information on the near side of disk and jet,
  which is available for 10 radio galaxies (see Section~\ref{s:nearfarside}). The thick solid curve indicates
  the cumulative distribution for the seven radio galaxies for which estimates of individual
  $\theta_{\rm DJ}$ are available (see Section~\ref{s:nearfarside}). The shape of the latter curve does not
  depend much on the choice of $\theta_{\rm DJ}$ for 3C 296 within its allowed range
  $30\deg \leq \theta_{\rm DJ} \leq 73\deg$ (we used the lower limit). {\bf
  Bottom:} same as top panel, but now for the set of eight dust lanes. The lanes are assumed to be systems
  which are viewed edge-on (see Section~\ref{s:lanejetintrinsic}). Correspondence between line style and
  model is similar to that in the top plot. The thick curve, present in the top plot, cannot be determined
  for dust lanes as $\epsilon$ and individual estimates of $\theta_{\rm DJ}$ are not available for lanes. The
  main conclusion is that jets of radio galaxies with dust lanes have on average smaller misalignment angles
  than those in radio galaxies with dust ellipses. See Section~\ref{s:lanejetintrinsic} for details.}
\label{f:cdfthetadj}
\end{centering}
\end{figure*}

 We explore model C, the truncated Gaussian, to test for the presence of a peak in the $\theta_{\rm DJ}$ distribution outside $\theta_{\rm DJ} = 45\deg$. The best fitted truncated Gaussian from the Maximum Likelihood analysis indicates that at least half of the radio jets make an angle of $\sim 35\deg$ or more with the symmetry axis of the dust disks (see Figure~\ref{f:cdfthetadj}). The confidence intervals on the free parameters $\mu$ and $\sigma$ are degenerate at large $\sigma$. The reason is the $\theta_{\rm DJ}$ distribution asymptotically approaches a uniform distribution regardless of the value of $\mu$. This precludes a direct interpretation of the confidence levels and hence we do not explore them.   

In summary, the jet misalignment angle distribution in galaxies with dust ellipses is consistent with having a peak around $\theta_{\rm DJ} \sim 45\deg$. The width of this peak is not well constrained. The limited data set cannot rule out a spherically random distribution of misalignment angles or a very narrow peak at more than $95\%$ confidence. These conclusions do
not depend critically on the assumed thickness or ellipticity. Regardless of the assumed model, typically at least half of the radio jets make an angle of $45\deg$ or more with the symmetry axis of the dust disks. 

\subsection{From axes to vectors}
\label{s:nearfarside}

In addition to the disk inclinations and dust-jet PA differences we
can estimate the near side of dust disk and jet with respect
to the observer in 11 of the radio galaxies (of which 10 are at $D<150\Mpc$). This provides extra
constraints on the distribution function of jet-disk misalignment
angles. The flux and morphology asymmetries between the radio jets on
both sides of the nucleus have been interpreted as being due to
relativistic motion of the radio-emitting particles in intrinsically
identical jets emerging on opposite sides from the nucleus (e.g.,
Laing \etal \cite{Lai99}). Thus the jet pointed nearest to the line of sight
(i.e., the 'main jet') appears brighter as the radio flux is either
Doppler boosted or less Doppler dimmed in comparison to the jet on the
other side of the nucleus.  The column density of stellar light
obscured by the dust disk along the line of sight is larger for the
near side of the disk than for the far side. Thus one expects a
difference in stellar surface brightness on both sides of the
disk. Such an effect is indeed seen: in many cases, the major axis of
the dust disk divides the disk into a more and a less obscured
half. We identify the side with a lower stellar surface brightness
with the near side of the disk. The surface brightness difference is
not clearly present for disks with a low ellipticity. This is in
qualitative agreement with the assumption that the disks are
intrinsically close to circular in which case a lower ellipticity
indicates a more nearly face-on disk for which the difference in the
obscured amount of stellar light decreases. For 11 galaxies the
observations reveal a clear radio flux asymmetry in the jets and a
dust obscuration asymmetry which allows us to estimate the near side
of disk and jet (cf.\ Table~\ref{t:fr1}).

With the addition of near and far side information for jet and disk,
the projections of the jet and dust axes in the plane of the sky
become projections of a jet and dust vector. The main jet is projected
against either the near side or far side of the dust disk (see
Table~\ref{t:fr1}). The absolute position angle difference between the
two vectors can vary between $0\deg$ and $180\deg$ instead of the
$0\deg<\Delta{\rm PA}_{\rm DJ}<90\deg$ for the dust and jet axis. We
repeated the Maximum Likelihood analysis using this extra information
for the ten galaxies at $D<150\Mpc$. The $\theta_A^{\rm obs}$ inferred for model A decreases by $\la 5\deg$, indicating a slightly more peaked distribution of misalignment angles compared to the analysis without
near side information (see Figure~\ref{f:confidence}, middle row). The inferred width of the peak at $\theta_{\rm DJ}=45\deg$ for model B stays the same.  
  
For seven of the 11 galaxies more detailed modelling of the jet
asymmetries at radio frequencies has been published (see Table~\ref{t:fr1} for references).  These studies
report not only the near side of the jet but also an estimate of the
viewing angle $\theta_{\rm JL}$ to the main jet (cf.\
Table~\ref{t:fr1}). Under the assumption of circular disks, this
viewing angle constrains the main jet to pass through either of the
two points on each 'jet-circle' at which the circle defined by $\theta_{\rm JL}$ intersects (cf.\ Section~\ref{s:lowerlimitsthetadjcircle}).  These
four points correspond to two different $\theta_{\rm DJ}$. For
one point the main jet is projected against the near side of the dust
disk while for the other point it is projected against the far side of
the dust disk. Given our estimate of the near and far side of the dust
disk, we can determine a unique $\theta_{\rm DJ}^{\rm radio}$ which is
listed in Table~\ref{t:fr1}. At least five of the seven misalignment angles are $\theta_{\rm DJ}^{\rm radio} \geq 40\deg$, confirming that such large misalignments occur frequently.
Special support for this comes from the galaxies with $\theta_{\rm
DJ}^{\rm radio}$ which are much larger than the minimally required
misalignment angle $\theta_{\rm DJ}^{\rm min}$. In fact, the
distribution of $\theta_{\rm DJ}^{\rm radio}$ agrees especially well
with the distribution inferred from models B and C (see
Figure~\ref{f:cdfthetadj}).

\section{Intrinsic orientation of dust lanes and radio jets}
\label{s:lanejetintrinsic}


We need to define a three-dimensional orientation for dust lanes to
constrain the jet misalignment angles in radio galaxies with dust
lanes. The analysis in Section~\ref{s:ellipsecircular} shows that the morphology of the dust ellipses is consistent with randomly oriented disks. The circumference of the dust lanes is too
irregular to assign an ellipticity. It seems that the dust lanes are all viewed rather close to edge-on: the ratio of shortest to longest linear scale in the dust lanes is
typically similar or smaller than the axis ratio of the most elongated
dust ellipses in the sample.
  
To obtain an upper limit on the average misalignment, we will assume that all dust lanes are exactly edge-on
systems. It
follows from Eq.~(\ref{e:thetamin}) that the minimally-required
misalignment angle $\theta_{\rm DJ}^{\rm min}$ is maximal for
$\theta_{\rm los}=90\deg$, i.e., an edge-on disk. Thus we expect to
first approximation that the Maximum Likelihood analysis will infer
an upper limit to the median of the distribution of $\theta_{\rm DJ}$ if all
lanes are assumed to be edge-on. We verified that this is indeed the
case for the specific distribution of $\Delta$PA$_{\rm DJ}$ for the
eight lanes at $D<150\Mpc$.

The upper limit to the median misalignment
angle for dust lane radio galaxies is smaller than the median angle for dust disk radio galaxies (cf.\ Figure~\ref{f:cdfthetadj}). This conclusion is reached by comparing models A, B and C. The inferred angle $\theta_B=41\deg$ suggests that the data are not consistent with a peak in the misalignment angle distribution at $\theta_{\rm DJ} =45\deg$. On the contrary, the $\theta_A = 28\deg$ and the typical misalignment angle $\theta_{\rm DJ} \sim 20\deg$ from model C indicate that the observations are most consistent with such small misalignment angles. For model A the $\theta_A$ inferred for the dust lanes falls outside the two-sided $\sim 95\%$ confidence region inferred for dust ellipses (cf.\ Figure~\ref{f:confidence}). The exact confidence level depends on the assumed thickness, ellipticity of the dust disks and whether additional assumptions on near side of the jets and disks are included. Similarly, for model B, the inferred $\theta_B$ falls outside the two-sided $\sim 68\%$ confidence region for dust ellipses. 

The fact that dust lanes seem to be viewed close to edge-on prompts the
question: where are the relatively face-on dust lanes? We explore several possibilities.
It could be that close to face-on dust lanes are not classified by us as 'lanes'. At $D<150\Mpc$ there are five galaxies with dust classified as intermediate between lane and disk. Three of these have axis ratios larger than 0.5 (i.e., roundish) and two have axis ratios less than 0.5 (i.e., flattened). Two of the three roundish galaxies have optical jets (3C 66B: Butcher \etal \cite{But80} and 3C 78: Sparks \etal \cite{Spa95}), suggesting the jet is pointed close to the line of sight (e.g., Sparks \etal \cite{Spa00}). Thus at least these two galaxies are consistent with having dust structures which are close to face-on and having jets roughly perpendicular to them, i.e., face-on counterparts of dust-lane systems with perpendicular jets. Furthermore, the two galaxies with axis ratios $<0.5$ have $\dpadj=68\deg$ and $\dpadj=76\deg$, i.e., the large angles typical for the dust lane galaxies. 

There are also five galaxies with dust classified as irregular. The dust in these galaxies typically shows several dusty filaments crossing the nucleus at several position angles. It is not implausible that these perturbed systems are the closer to face-on counterparts of perturbed dust lanes. The dust size distribution of the combined subset of galaxies with either intermediate dust or irregular dust is similar to that of dust lanes (see Figure~\ref{f:dpa} and Table~\ref{t:fr1}). In one galaxy with irregular dust (M87) an optical jet is seen. In fact,    
three of the four galaxies at $D<150\Mpc$ with optical jets in our sample are found in galaxies with dust classified as intermediate or irregular. The one other optical jet is in a dust disk galaxy (NGC 3862). This galaxy has a very round dust structure (Table~\ref{t:fr1}). Thus one can speculate that this might also be a 'face-on dust lane' galaxy where the jet is perpendicular to the disk (see also Sparks \etal \cite{Spa00}). It is interesting to note that two other round dust ellipses (NGC 541 and UGC 7115) do not have optical jets. Thus these could then be the dust disk galaxies for which we inferred that the jets are not perpendicular to the disks. In conclusion there is evidence that many of the ten galaxies which contain intermediate or irregular dust could be the closer to face-on counterparts of the eight rather edge-on dust lane galaxies. 

\section{Discussion}
\label{s:discussion}

We repeat the three questions posed in the Introduction which form the focus of our analysis.

{\sl Are the properties of the fuel reservoirs different in radio galaxies and
non-radio galaxies?} 

In terms of dust morphology the answer is yes: non-radio galaxies more often display highly irregular dust distributions. Perhaps this reflects the fact that the irregular fuel reservoirs are falling into the nucleus of the galaxy but are not feeding the black hole yet to cause nuclear activity and the formation of jets. In terms of reservoir orientation the answer is no. The orientations of dusty filaments and disks have similar distributions with respect to the optical isophotes for radio and non-radio galaxies. 


{\sl Is the dust orientation governed by the
gravitational potential?} 


Giant ellipticals are generally thought to be stationary or at most slowly rotating triaxial structures. Such galaxies
have two equilibrium planes for dust disks, i.e., planes where stable
non-intersecting orbits exist (e.g., Merritt \& de Zeeuw \cite{Mer83}). These
planes are perpendicular to the short and long axis respectively. The
angles between the minor axis and the short and long axis,
respectively, depend on the line of sight and triaxiality
$T=(1-p^2)/(1-q^2)$ of the host galaxy (e.g., Franx \etal \cite{Fra91}). If we assume that ellipses and
lanes are both intrinsically circular structures, then their minor
axis and hence $\Delta{\rm PA}_{\rm DG}$ constrain the rotation axis assuming that the dust is settled.

\begin{figure*}[t]
\begin{centering}
\resizebox{1.0\hsize}{!}{\includegraphics{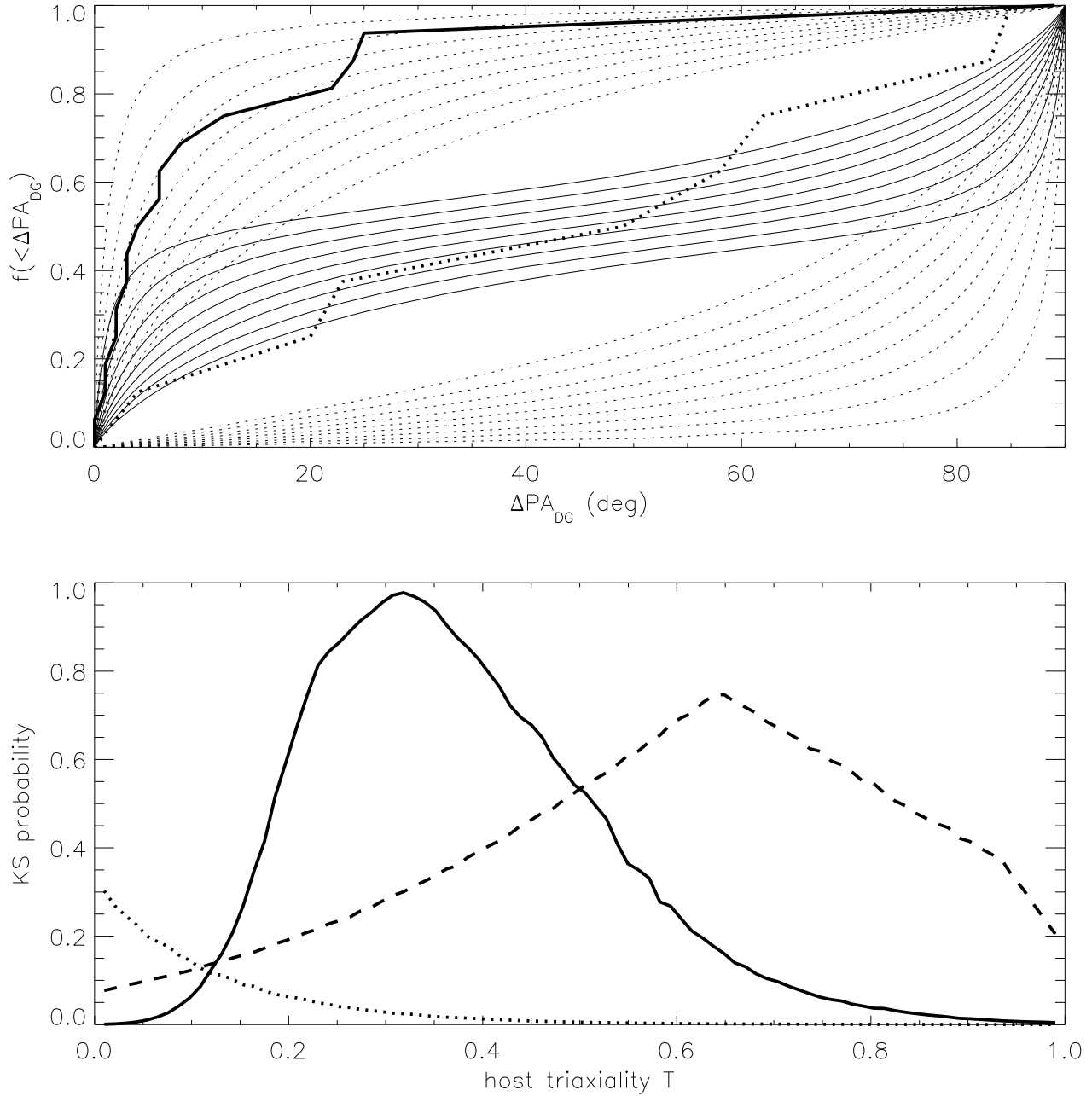}}
\caption{{\bf Top:} the cumulative distribution of
  $\Delta{\rm PA}_{\rm DG}$. The two thick lines show the observations
  for the FR sample galaxies at $D<150\Mpc$ for dust ellipses (solid
  line) and dust lanes (dotted line). The thin dotted lines show the
  expected cumulative distribution of $\Delta{\rm PA}_{\rm DG}$ for
  circular disks rotating around the short axis (top series) and
  around the long axis (bottom series) assuming random viewing
  angles. The thin solid lines are the combination of these two
  distributions. They indicate the cumulative distribution for an
  equal number of disks rotating around the short and long axis. The
  nine curves in each series indicate the expectation for a stellar
  host triaxiality of 0.1 (top) increasing to 0.9 (bottom) in steps of
  0.1. {\bf Bottom:} the
  Kolmogorov-Smirnov (KS) probabilities that the observed
  $\dpadg$ distributions in the top plot are drawn from modelled
  distributions as a function of the stellar host triaxiality $T$.  The computation assumes that the dust is a thin circular
  disk at random viewing angles and takes into account observational
  errors. The KS probability is significant at certain $T$ only for
  dust ellipses rotating around the short axis (solid line), lanes
  rotating around the long axis (dotted line) and lanes rotating in
  equal numbers around the short and long axis (dashed line). Figure~\ref{f:triaxiallane} shows a similar analysis for dust lanes assuming viewing angles which are closer to edge-on. See
  Section~\ref{s:discussion} for further details.}
\label{f:triaxial}
\end{centering}
\end{figure*}

\begin{figure*}[t]
\begin{centering}
\resizebox{1.0\hsize}{!}{\includegraphics{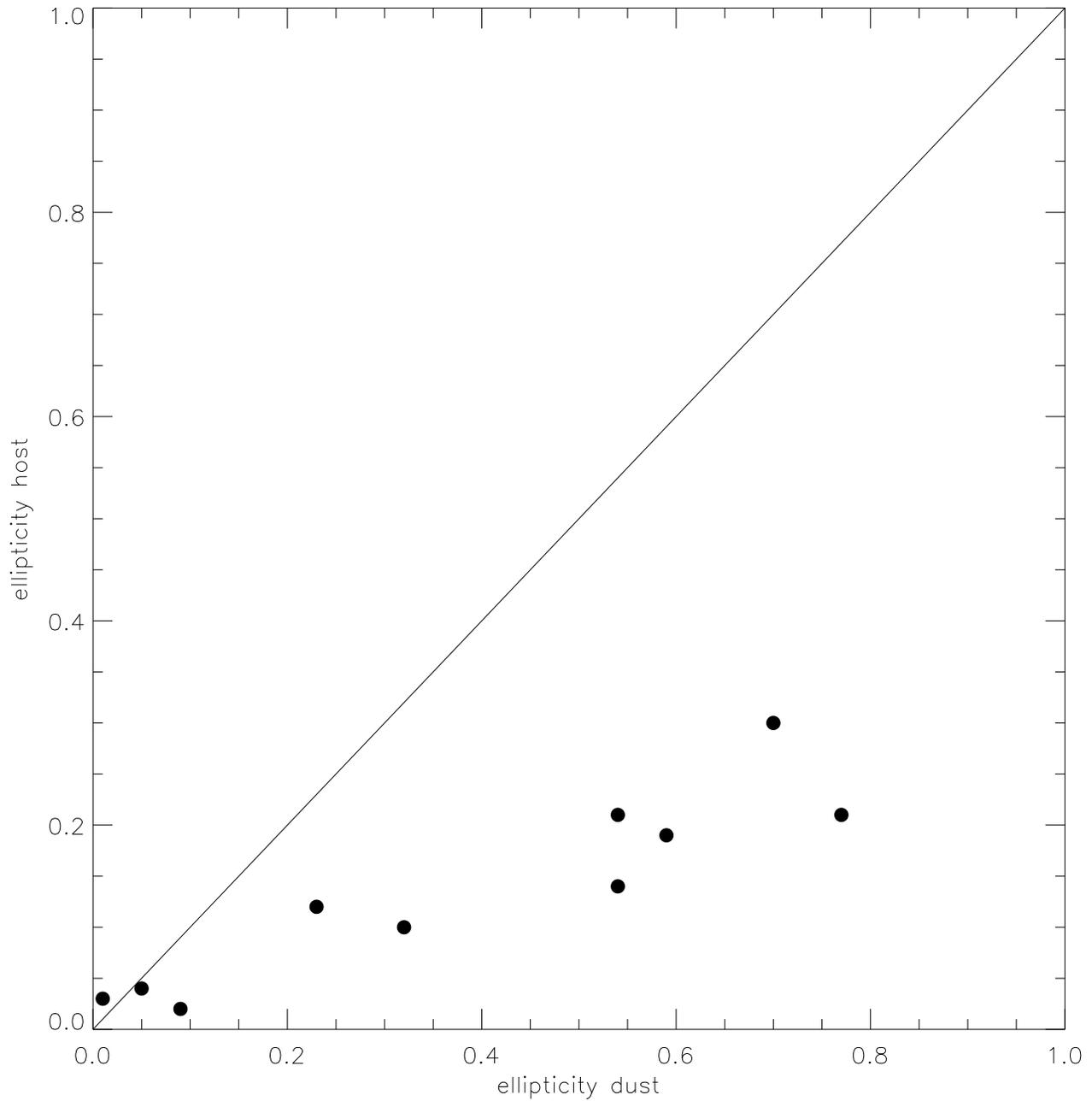}}
\caption{The ellipticity of the stellar host as a 
  function of dust ellipticity for the dust disks in the UGC~FR-I
  sample. The host ellipticity is taken from HST imaging at a radius
  just outside the main dust distribution, using the results in Verdoes
  Kleijn \etal (\cite{Ver99}). As expected for settled dust rotating around the short axis, the dust ellipses
  have ellipticities which are (i) systematically larger than host
  ellipticities and (ii) increasing for increasing host
  ellipticity.}
\label{f:epsdepsg}
\end{centering}
\end{figure*}

The top panel in Figure~\ref{f:triaxial} shows for the FR sample
galaxies (at $D<150\Mpc$) the observed cumulative distribution
functions of $\Delta{\rm PA}_{\rm DG}$ for dust ellipses and those predicted for randomly-oriented thin
circular disks rotating around the short and/or long axis as a
function of host triaxiality. The bottom panel shows the corresponding
probabilities from a Kolmogorov-Smirnov test that the observed
distribution is drawn from a model distribution with a specific
stellar host triaxiality. In computing the KS probabilities we added
typical errors to the model distributions. The errors have a skewed
distribution because $0\deg < \Delta{\rm PA}_{\rm DG} < 90\deg$ by
definition. Taking this into account significantly influences the
probability distribution.  The $\dpadg$ of ellipses are only
consistent with disks rotating around the short axis. The preferred
host triaxiality $T \sim 0.30$ is close to oblate, in agreement with
early studies of giant ellipticals (e.g., Franx et al.\ \cite{Fra91}). In this
scenario of settled disks, one would expect to first approximation (i)
dust ellipticities which are always larger than stellar host
ellipticities and (ii) increasing host ellipticity for increasing disk
ellipticity. This is exactly what is observed (Fig.~\ref{f:epsdepsg}).

Examining the properties of individual galaxies instead of general sample properties , we note two radio
galaxies which could be exceptions to the above scenario: NGC 4261 and 3C 76.1. NGC 4261 is an unusual galaxy
because the stellar mean motion is around the major axis (Davies \& Birkinshaw \cite{Bir85}). This could
indicate that the galaxy is prolate. In that case the dust disk would lie in a plane which is not
perpendicular to the long-axis. Such a plane is dynamically unstable and one would expect that the gas would
have settled down in another configuration. Thus it is more likely that the galaxy is triaxial but close to
prolate. In that case the orientation of the dust disk and mean rotation axis of the stars are consistent
with the picture in which the dust disk lies in a plane perpendicular to the short axis of the galaxy, which
is stable. The second stable plane in a triaxial galaxy, perpendicular to the long axis, is more unlikely
given that the disk is elongated in the direction of the galaxy major axis. 3C 76.1 contains the only dust ellipse in
the sample which is aligned with the minor axis of the galaxy instead of the major axis. Thus, this galaxy
could be prolate with a settled dust disk in the stable plane perpendicular to the long-axis. Thus two out of
18 galaxies might be (nearly) prolate instead of (nearly) oblate. This fraction is consistent with the
observed distribution of stellar rotation axes in large samples of ellipticals (Franx \etal \cite{Fra91}).

\begin{figure*}[t]
\begin{centering}
\resizebox{1.0\hsize}{!}{\includegraphics{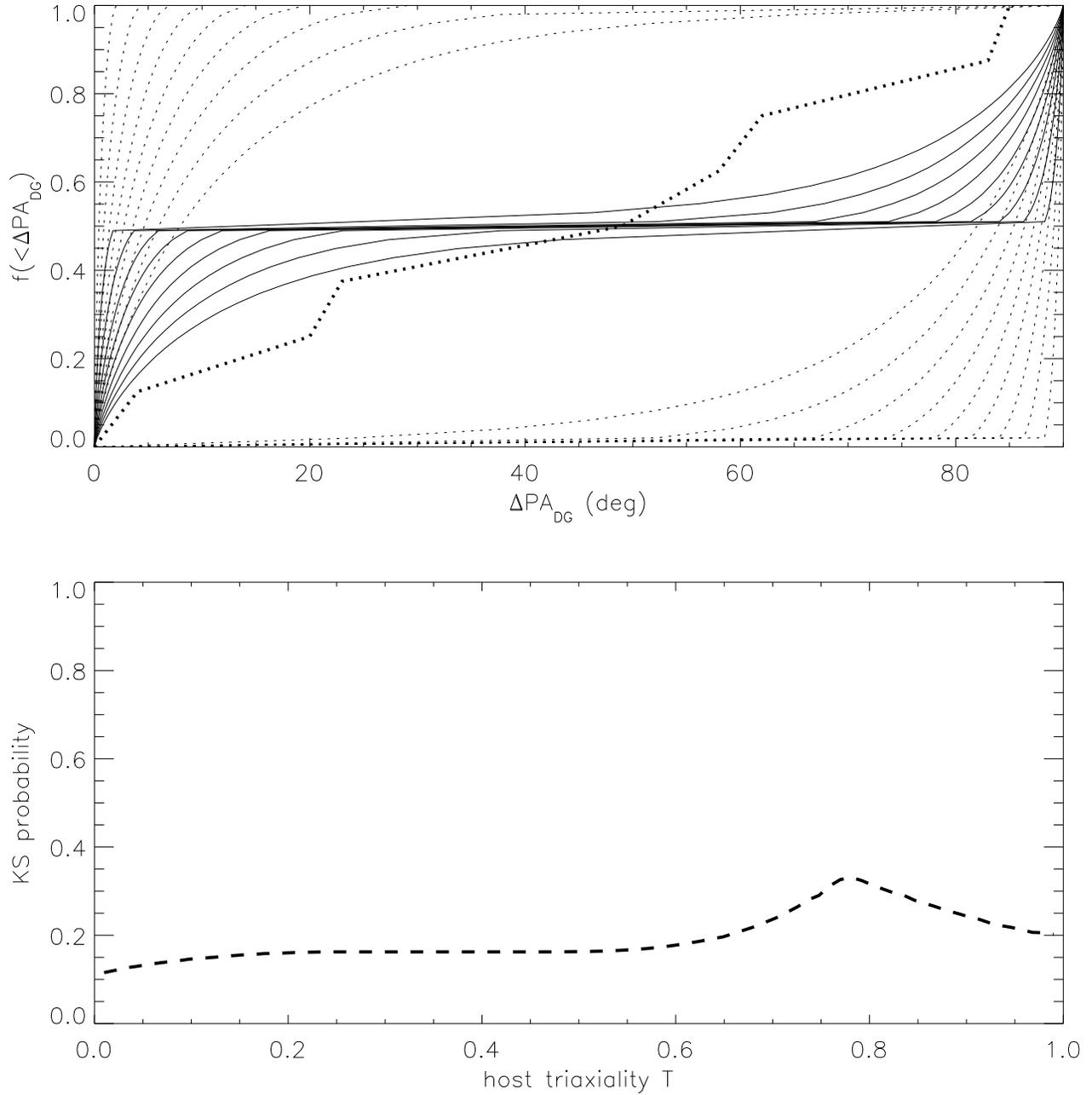}}
\caption{Similar plots as Figure~\ref{f:triaxial}, but now for a model in which the thin
  circular disks are viewed within $30\deg$ from edge-on instead of fully random viewing angles. This model
  is more plausible for dust lanes. The Kolmogorov-Smirnov probabilities for lanes are only significant for
  lanes rotating around the short and long axis of the triaxial host. Nevertheless, this model does not
  provide a good fit to the data for any triaxiality $T$. In other words, it appears unlikely that the dust
  lanes are consistent with being settled structures in any of the two equilibrium planes in a triaxial host,
  but the data set is too small to rule out such a scenario completely.}
\label{f:triaxiallane}
\end{centering}
\end{figure*}

To constrain the host shape for dust lane galaxies we start out with assuming that also dust lanes are thin
circular disks viewed at random viewing angles.  Figure~\ref{f:triaxial} shows that the $\dpadg$ of dust lane
galaxies is most consistent with equal numbers of dust lanes rotating around the short- and long axis in
highly triaxial ($T \sim 0.65$) galaxies. Such a high triaxiality appears inconsistent with other studies of
the shapes of giant ellipticals (e.g., Franx \etal \cite{Fra91}). Furthermore, the similarity in large-scale axis ratios, optical luminosities and central stellar velocity dispersion suggests that dust lanes and dust ellipses live in similar hosts (Figure~\ref{f:ugcnormalsample}). Figure~\ref{f:triaxial} shows
that the data cannot rule out at high confidence that lanes occupy systems with a
typical triaxiality $T \sim 0.2-0.3$ but with lanes rotating around the long axis or around the short and
long axis. However, as discussed in Section~\ref{s:lanejetintrinsic}, the assumption of random viewing angles
appears implausible as dust lanes are more likely to be viewed relatively close to edge-on. Thus, we repeat
the KS analysis but now we constrain the viewing angles to lie within $30\deg$ from edge-on. The result is
shown in Figure~\ref{f:triaxiallane}. The hypothesis that all lanes are dust structures settled in the
equilibrium plane perpendicular to either the short or the long axis is ruled out with $>95\%$
confidence. A model in which the dust lanes rotate in equal numbers around both axes cannot be ruled out at
high confidence, but appears to be a poor fit to the observed $\dpadg$ as well.

The simultaneous differences between lanes and ellipses in both their $\dpadg$ distribution and morphology  support the alternative idea that the lanes are not simply settled, almost edge-on disks of dust. It is more likely that the dust lanes are
perturbed disks. Models of a warped disk have been used to
explain the filamentary dust morphology in NGC 4374 (M84; Quillen \&
Bower \cite{Qui99}) and both the morphology and kinematics of the interstellar
matter in NGC 5128 (Quillen \etal \cite{Qui92}). For NGC 4374 and NGC 5128
the studies infer an angle between the line of sight and the symmetry
axis of the warped disk of $68\deg$ and $60\deg$ respectively. In
general, the perturbations and warps of the disks classified as
'lanes' could be non-transient if they are caused by triaxiality of the gravitational potential or
constantly perturbing forces related to the active galactic nuclei and jets.
The perturbations
would be transient if the dust lanes are in the process of settling
down to become flat circular disks (Steiman-Cameron \etal \cite{Ste92}), at which point we would classify them as dust
ellipses. There is the following qualitative support
for this. Some disks have in addition more irregular dust at
larger scales (e.g., NGC 315, NGC 4335 and NGC 4261). In particular
the case of NGC 4335 is interesting as there the larger scale dust is
attached to the disk and has an arc-like morphology very reminiscent
of a settling disk (e.g., Verdoes Kleijn \etal \cite{Ver02}). Some of
the most regular lanes have small $\dpadg$ (e.g., NGC 5127 and NGC
5141). Furthermore, (some of) the five radio galaxies at $D<150\Mpc$ with intermediate dust structures might be literally in between lane and ellipse. All have dust sizes $<550\pc$ which is typical for the settled dust ellipses, but their dust orientation is not aligned with the galaxy major axis. For the nine  cases at $D > 150\Mpc$ the classification might be better described as 'ambiguous' due to large distance instead of intermediate (see Section~\ref{s:ellipsecircular}).  
Lastly, the hosts of dust ellipses seem close to oblate in general. In
oblate galaxies the dynamical settling times are largest for lanes
perpendicular to the equatorial plane. At the same time, the settling
process is most likely dissipational such that the dust structures become
more compact as settling proceeds. Thus one expects larger lanes to have on average larger $\dpadg$. This is observed (see
Fig.~\ref{f:dustcomparison}). Similarly, one expects large dust distributions to display closer alignment with the major axis at smaller radii because of the decreased dynamical time scale. NGC 4335 displays this behaviour, but it is not seen in general. 

It is reasonable to assume that the
current dust distribution was accreted by the galaxy before or around
the time of the on-set of the AGN. 
The typical ages of FR I radio jets estimated from
dynamical arguments is $\sim 10^8$yr and from radiative arguments
$\sim 10^7$yr (e.g., Eilek \etal \cite{Eil97}). Settling time-scales for the
accreted gas are typically estimated to be a few orbital time-scales
(e.g., Steiman-Cameron \cite{Ste91}). Orbital times are typically $\sim 2
\times 10^6 (r/100{\rm pc})$ yr for giant ellipticals, where $r$ is
the radius for a circular orbit. (This estimate assumes that the
orbital circular velocities are on the order of the central velocity
dispersions as measured in the central few arcsec$^2$.) The lifetimes
of the AGNs seem a bit long for the dust still to be settling
down. However, the uncertainties in both time-scales are too large to
make firmer statements.
 


{\sl Do jet/AGN related forces affect the dust
orientation?} 


The dust/galaxy alignment dichotomy holds for giant elliptical galaxies in general, might they be radio or non-radio galaxies. A main result of our work is that for radio galaxies the dust/galaxy alignment dichotomy is connected to a jet misalignment dichotomy in three dimensions. Filamentary dust structures, i.e., the dust lanes, tend to be perpendicular to jets and to have an arbitrary orientation with respect to the galaxy major axis. Well-defined elliptical dust structures, i.e., the dust ellipses, have an arbitrary orientation with respect to the radio jets and tend to be aligned with the galaxy major axis. We explore qualitatively two scenarios to explain the dust/galaxy/jet orientation dichotomy.

Scenario I: the jets produce a torque in the hot X-ray gas. The torque is such that the X-ray gas tries to force the cooler dusty ISM in a plane perpendicular to the jet. Such a model was explored in detail for NGC 4374 by Quillen \& Bower (\cite{Qui99}). The stellar mass potential produces a torque which tries to force the ISM in the stable settling planes of the galaxy potential. If the X-ray gas torque is stronger than the gravitational torque, the dust is forced in a plane perpendicular to the jet. This would produce the dust lanes galaxies. If the X-ray gas torque cannot overcome the gravitational torque the dust settles on non-intersecting orbits in the settling plane. This would produce the dust disk galaxies. It seems not implausible that an X-ray torque which is similar or stronger than the gravitational torque results in a perturbed disk with warps and filaments explaining the more irregular appearance of dust lanes compared to the dust ellipses (Quillen \& Bower \cite{Qui99}). This scenario does not explain why the dust/galaxy misalignment is also seen in galaxies without radio jets. However, it is interesting to note that none of the dust lanes in non-radio galaxies have $20\deg \la \dpadg < 80\deg$. In particular, small ($\la \kpc$) dust lanes are oriented close to the galaxy major axis and all large ($\ga \kpc$) dust lanes close to the minor axis. It might be that the small dust lanes in non-radio galaxies can be explained by dust structures settled in planes perpendicular to the short axis in a mildly triaxial host. The large-scale dust lanes in non-radio galaxies might be still settling down towards this plane. In contrast to the non-radio galaxies, quite a few radio galaxies have misalignments at $\dpadg \sim 45\deg$ and quite a few of those have small sizes. This difference could indicate that dust lanes in radio galaxies are prevented from settling down by an additional force. 

Scenario II: dust is acquired externally (or from the outskirts of the galaxy) and the orientation of jet is aligned with the initial angular momentum of the dusty in-falling material. In this scenario the jet direction is determined during its formation by the angular momentum of in-falling material. A speculative example: it could be that an in-falling small galaxy brings in not only the dust and gas that we see, but also a central black hole. Initially the orientation of the dusty material then indicates the plane in which this black hole orbits the supermassive black hole of the giant elliptical. The angular momentum of the subsequent black hole binary or merger determines the orientation of the jets. Thus, initially the dust is expected to be perpendicular to the jet and to have a perturbed morphology because it has not settled yet. As time goes on the collisional dust and gas settles down in the galaxy potential thereby loosing its relation with the jet orientation. This scenario explains the dust/galaxy/jet dichotomy and the related differences in dust morphology. It also naturally explains why the dust/galaxy dichotomy is seen both in radio and non-radio galaxies. A weakness of the model is that the settling time-scales of the dust seem short compared to the age of the radio sources.

\section{Conclusions}
\label{s:conclusions}

The main conclusions from the analysis of dust in 21 nearby FR-I radio galaxies and a comparison sample of 52 non-radio giant ellipticals are:
\begin{itemize}
\item{The detection rate of dust is higher in radio galaxies ($\sim 90\%$) than in non-radio galaxies ($\sim 50\%$).}
\item{The dust morphology can be classified as disk-like ellipses, filamentary lanes and irregular.}
\item{Non-radio galaxies contain more often irregular dust distributions.}
\item{The dust morphology, size, orientation relative to the stellar host of ellipses and lanes is similar in radio and non-radio galaxies.}
\item{Filamentary dust, i.e., lanes, are generally misaligned with the galaxy major axis. Dust with a smooth elliptical appearance, i.e., dust ellipses, are roughly aligned with the galaxy major axis.}
\item{The dust lanes and ellipses live in similar hosts in terms of stellar luminosity, axis ratios and velocity dispersions.}
\end{itemize}

The main conclusions from the analysis of the relative orientation (in two and three dimensions) of dust, host galaxy and radio jet in a sample of 47 radio galaxies are:
\begin{itemize}
\item{Dust ellipses are consistent with being relatively thin, circular or slightly elliptical disks rotating around the short axis of an oblate-triaxial gravitational potential with a triaxiality $T \sim 0.3$.}
\item{The dust lanes are most likely warped or perturbed dust structures viewed rather close to edge-on. However, a scenario in which the dust lanes are settled structures in any of the two equilibrium planes of an oblate-triaxial galaxy cannot be ruled out completely.}
\item{The radio galaxies with dust with an apparent morphology classified as either irregular or intermediate between ellipse and lane most likely represent the close to face-on counterparts of the edge-on lanes.}
\item{The radio jets do not have a particular orientation with respect to the dust disks. The mean angle between the disk normal and jet (the 'misalignment angle') is inferred to be $\sim 45\deg$.}
\item{The radio jets tend to be perpendicular to the dust lanes, both in two and three dimensions. The upper limit to the mean misalignment angle is $\sim 20\deg-30\deg$.}
\item{The above two results on jet orientations do not depend on the allowed range of ellipticity and thickness of the dust structures.}
\item{The use of independent information on the near side of jets and dust disks strengthen the above results on jet/dust orientation.}
\end{itemize}

There are two scenarios to explain the the dust/galaxy/jet orientation dichotomy for radio galaxies. 
Scenario I: the radio jets exert a torque on the nuclear dust in active galaxies which in some cases overcomes the gravitational torque causing the nuclear dust in a plane perpendicular to the jets.
Scenario II: the angular momentum vector of unsettled nuclear dust is initially aligned with the radio jet in active galaxies but this alignment is lost as dust settles in an equilibrium plane of the galaxy.

To distinguish between these scenarios we deem the following steps relevant. First, larger samples of dusty galaxies, both active and inactive are required to confirm the dichotomy in jet/dust/galaxy orientation and its details. The current confidence levels on the existence of the dichotomy are $\la 95\%$. Second, more detailed analysis of host galaxy properties, such as gravitational potential and X-ray gas pressure gradients, are required to obtain better estimates of the settling and life time of the dusty ISM. Third, combining such an analysis with detailed studies of the dusty and gaseous ISM, such as metallicity and two-dimensional kinematics, one might hope to constrain the origin of the dusty fuel reservoirs: stellar mass loss, cooling flows, or accretion of satellites. 


\begin{acknowledgements}
Support for this work was provided by NASA through grant number
\#GO-06673.01-95A from the Space Telescope Science Institute, which is
operated by AURA, Inc., under NASA contract NAS5-26555. The authors
thank Stefi Baum and Henrique Schmitt for discussions that helped to improve
the manuscript, and Anne-Marie Weijmans for help in testing the software used in Sections~\ref{s:ellipsejetintrinsic} and ~\ref{s:lanejetintrinsic}. The authors are grateful to the referee for his/her comments. They helped to improve significantly the presentation of the results.
\end{acknowledgements}



\ifsubmode\else
\baselineskip=10pt
\fi


\clearpage


\clearpage
 
\clearpage

\clearpage
 
\clearpage

\clearpage

\clearpage

\clearpage

\clearpage
 
\clearpage
 
\clearpage
  
\clearpage



\clearpage

\begin{longtable}{lccrcrrrrcrrl}
\caption{\label{t:fr1} Dust properties for the FR sample of radio galaxies. Properties of dust detected in the FR sample, which
includes the 21 UGC~FR-I radio galaxies as a subset (see
Section~\ref{s:samples}). The latter are listed above the double horizontal
lines. No dust is detected in two UGC~FR-I galaxies: NGC 741 and NGC
2892. Col.(1): target name. A $^*$ indicates radio sources which are
sometimes classified as FR-Is (e.g., de Koff \etal \cite{dKof00}) and sometimes
as FR-IIs (e.g., Chiaberge \etal \cite{Chi02}) or as intermediate, i.e., FR-I/II.  Col.(2): distance to the
galaxy taken from Faber \etal \cite{Fab89} or LEDA, except for NGC 5128 (Israel \cite{Isr98}). Col.(3): dust morphology classification (see
Section~\ref{s:dustproperties} for their definition). 1= dust lane, 2=
dust ellipse, 3= irregular dust. A 1-2 classification indicates an
intermediate morphology classification: they could be either lanes or
ellipses. Col.(4): longest linear extent
of the dust which has a typical relative error of $10\%$. Entries
marked with a $^1$ have an additional irregular extended
component. Col.(5): ellipticity for dust ellipses. The error is given
in between the brackets. We measured the ellipticity $\epsilon$ from
the HST imaging except for 3C402N, which was taken from Schmitt \etal
(\cite{Sch02}). Col.(6): dust position angle with its error in between brackets. This is the PA of the major axis
for dust ellipses and of the longest dust extent for dust lanes. An
$^1$ indicates that the dust feature was too faint to measure its
position angle reliably.  Col.(7): the position angle difference
between the dust and galaxy major axis. The error is listed in between
brackets. A $^2$ indicates that the galaxy is too round to determine
its position angle. Col.(8): position angle difference between jet
axis and dust ellipse major axis or longest axis of the dust lane with
the error in between brackets. Col.(9): minimum misalignment angles
$\theta^{\rm min}_{\rm DJ}$ between the jet axis and the normal of the
dust disk. The first and second value are obtained assuming thin ($q=0$)
circular disks ($p=1$) and elliptic disks ($p=0.75$),
respectively. Col.(10): side of dust disk against which the main jet is projected;
near/far indicates the side nearest or farthest from the observer,
respectively. See Section~\ref{s:nearfarside} for details on its
determination. Col.(11): viewing angle to the main jet (i.e., the
angle between the line of sight and the jet) as inferred from jet
asymmetries at radio frequencies. This angle is taken from the
literature (see references). Col.(12): The jet misalignment angle as
inferred from the viewing angle to the main jet. Col.(13): reference
for PA of jet and galaxy. 1: de Koff \etal (\cite{dKof00}); 2: Martel \etal
(\cite{Mar99}); 3: Capetti \etal (\cite{Cap00}); 4: Schmitt \etal (\cite{Sch02}); 5: de Koff
\etal (\cite{dKof96}); 6: de Juan \etal (\cite{dJua96}); 7: Guthrie (\cite{Gut80}) (we measured
the jet PA from Fig.~1); 8: Dufour \etal (\cite{Duf79}); 9: Schreier \etal
(\cite{Sch81}); 10: VIMOS imaging (unpublished); 11: Verdoes Kleijn \etal
(\cite{Ver99}); 12: Verdoes Kleijn \etal (\cite{Ver02}); 13: Leahy \& Perley (\cite{Lea91}). References for $\theta_{\rm
JL}$: 13: Laing \etal (\cite{Lai04}); 14: Laing \etal (\cite{Lai99}); 15: Feretti
\etal (\cite{Fer99}); 16: Hardcastle \etal (\cite{Har97})
}\\
\hline\hline
target & D & morph & size & $\epsilon$ & PA$_{\rm D}$ & $\Delta$PA$_{\rm DG}$ & $\Delta$PA$_{\rm DJ}$ & $\theta^{\rm min}_{DJ}$ & side & $\theta_{\rm JL}$ & $\theta_{\rm DJ}^{\rm radio}$ & ref \\
 & (Mpc) & & (pc) & & ($\deg$) & ($\deg$) & ($\deg$) & ($\deg$) & & ($\deg$) & ($\deg$) & \\
\hline
\endfirsthead
\caption{continued.}\\
\hline\hline
target & D & morph & size & $\epsilon$ & PA$_{\rm D}$ & $\Delta$PA$_{\rm DG}$ & $\Delta$PA$_{\rm DJ}$ & $\theta^{\rm min}_{DJ}$ & side & $\theta_{\rm JL}$ & $\theta_{\rm DJ}^{\rm radio}$ & ref \\
 & (Mpc) & & (pc) & & ($\deg$) & ($\deg$) & ($\deg$) & ($\deg$) & & ($\deg$) & ($\deg$) & \\
\hline
\endhead
\hline
\endfoot
NGC 193    &  58 & 1   &  840$^{1}$              & ...        &   0(6)   & 58(7)  & 77(10) & ...  & ...  & ...  & ... & 11 \\
NGC 315    &  68 & 2   &  830$^{1}$              & 0.77(0.02) &  40(2)   &  1(4)  & 90(8)  &  0/0 & far  & 37   & 40  & 11,13 \\ 
NGC 383    &  65 & 2   & 2350$^{\hskip 0.13cm }$ & 0.23(0.03) & 138(2)   &  6(3)  & 24(8)  & 36/0 & far  & 52   & 47 & 11,13 \\ 
NGC 541    &  73 & 2   &  640$^{\hskip 0.13cm }$ & 0.09(0.05) & ...      &   ...  & ...    & ...  & ...  & ...  & ... & 11 \\
UGC 1841   &  85 & 1-2 &  330$^{\hskip 0.13cm }$ & $<0.5$     & ...      &   ...  & ...    & ...  & ...  & ...  & ... & 11 \\
NGC 2329   &  77 & 2   &  750$^{\hskip 0.13cm }$ & 0.32(0.05) & 174(3)   &  3(4)  & 24(9)  & 42/29& near & ...  & ... & 11 \\ 
NGC 3801   &  43 & 1   & 4430$^{1}$              & ...        &  24(2)   & 83(4)  & 89(8)  & ...  & ...  & ...  & ... & 11 \\ 
NGC 3862   &  84 & 2   &  610$^{\hskip 0.13cm }$ & 0.01(0.02) & ...      &   ...  & ...    & ...  & ...  & ...  & ... & 11 \\
UGC 7115   &  91 & 2   &  570$^{\hskip 0.13cm }$ & 0.05(0.03) & ...      &   ...  & ...    & ...  & ...  & ...  & ... & 12 \\
NGC 4261   &  30 & 2   &  240$^{1}$              & 0.54(0.03) & 163(1)   &  8(5)  & 76(8)  & 12/4 & far  & ...  & ... & 11 \\ 
NGC 4335   &  62 & 2   &  750$^{1}$              & 0.59(0.10) & 158(2)   &  2(5)  & 79(8)  & 10/3 & ...  & ...  & ... & 11 \\
NGC 4374   &  15 & 1   &  990$^{\hskip 0.13cm }$ & ...        &  79(10)  & 49(11) & 80(13) & ...  & ...  & ...  & .... & 11 \\
NGC 4486   &  15 & 3   &  560$^{\hskip 0.13cm }$ & ...        & ...      &   ...  & ...    & ...  & ...  & ...  & ... & 11 \\
NGC 5127   &  64 & 1   &  950$^{\hskip 0.13cm }$ & ...        &  48(2)   & 20(5)  & 70(8)  & ...  & ...  & ...  & ... & 11 \\
NGC 5141   &  71 & 1   &  550$^{\hskip 0.13cm }$ & ...        &  88(8)   & 23(11) & 76(11) & ...  & ...  & ...  & ... & 11 \\
NGC 5490   &  70 & 1-2 &  170$^{\hskip 0.13cm }$ & $\ga 0.5$ & 143(4)   & 38(6)  & 68(9)  & ...  & ...  & ...  & ... & 11 \\
NGC 7052   &  55 & 2   & 1080$^{\hskip 0.13cm }$ & 0.70(0.02) &  65(1)   &  1(1)  & 42(8)  & 45/40& far  & 79   & 47  & 11,14 \\
UGC 12064  &  68 & 2   & 1200$^{\hskip 0.13cm }$ & 0.54(0.05) & 166(5)   &  6(5)  & 25(9)  & 54/43& near & 83   & 72  & 12,15 \\
NGC 7626   &  47 & 1   &  230$^{\hskip 0.13cm }$ & ...        & 175(4)   &  4(6)  & 57(9)  & ...  & ...  & ...  & ... & 11 \\
\hline
\hline
NGC 5128   &  3.4& 1   & 9200$^{\hskip 0.13cm }$ & $>0.5$     & 120(3)   & 85(4)  &  67(9) & ...  & ...     & ...     & ... & 8,9,10\\
3C15$^*$   & 292 & 1-2 &  985$^{\hskip 0.13cm }$ & $<0.5$     & $^1$     & ...    &  ...   & ...  & ...     & ...     & ... & 2 \\
3C29       & 180 & 1-2 &  486$^{\hskip 0.13cm }$ & $<0.5$     & $^1$     & ...    &  ...   & ...  & ...     & ...     & ... & 2 \\
3C40$^*$   &  72 & 2   &  146$^{\hskip 0.13cm }$ & 0.53(0.07) &  77(7)& 25(8)  &  62(11)& 24/17& far     &         &     & 2 \\
3C75       &  96 & 1-2 &  194$^{\hskip 0.13cm }$ & $> 0.5$    &  36(5)& 65(5)  &  76(9) & ...  & ...     & ...     & ... & 2 \\
3C76.1     & 128 & 2   &  892$^{\hskip 0.13cm }$ & 0.65(0.05) &  39(4)& 89(5)  &  72(9) & 17/11& ...     & ...     & ... & 13 \\
3C78       & 116 & 1-2 &  469$^{\hskip 0.13cm }$ & $< 0.5$    & $^1$  & ...    &  ...   & ...  & ...     & ...     & ... & 2 \\
3C83.1     & 102 & 2   & 1020$^{\hskip 0.13cm }$ & 0.86(0.03) & 168(2)   &  2(4)  & 82(8)  &  8/6 & ...     & ...     & ... & 1,2\\
3C84       &  72 & 3   &12060$^{1}$              & ...        & ...      &    ... &    ... & ...  & ...     & ...     & ... & 2\\
3C88$^*$   & 121 & 1-2 & 550$^{\hskip 0.13cm }$  & $\la 0.5$ & $^1$     &    ... & ...    & ...  & ...     & ...     & ... & 1,2\\
3C296      &  95 & 2   &  568$^{\hskip 0.13cm }$ & 0.71(0.05) & 157(4)   &  12(4) & 60(9)  & 29/23& far     &$\leq 63$& $30-73$& 1,2,16\\
3C315      & 433 & 1-2 & 1480$^{\hskip 0.13cm }$ & $\ga 0.5$ & $^1$     &    ... &    ... & ...  & ...     & ...     & ... & 1,5\\
3C317      & 140 & 3   &11170$^{\hskip 0.13cm }$ & ...        & ...      &    ... &    ... & ...  & ...     & ...     & ... & 2\\
3C338      & 119 & 3   & 1750$^{\hskip 0.13cm }$ & ...        & ...      &    ... &    ... & ...  & ...     & ...     & ... & 1,2\\
3C353$^*$  & 122 & 1   & 1460$^{\hskip 0.13cm }$ & ...        & 148(7)   &    $^2$& 64(11) & ...  & ...     & ...     & ... & 1,2\\
3C402N     & 101 & 2   & 1080$^{\hskip 0.13cm }$ & 0.63(0.05) &  58(3)   &   3(5) & 63(9)  & 25/19& ...     & ...     & ... & 4,6\\
3C442$^*$  & 108 & 3   &  826$^{\hskip 0.13cm }$ & ...        & ...      &    ... &    ... & ...  & ...     & ...     & ... & 2\\
3C452$^*$  & 324 & 1-2 & 3060$^{\hskip 0.13cm }$ & $>0.5$     &   0(5)   &  79(6) & 79(9)  & ...  & ...     & ...     & ... & 2\\
4C-03.43   & 207 & 1-2 &  450$^{\hskip 0.13cm }$ & $\ga 0.5$ &  86(5)   &  68(6) & 69(9)  & ...  & ...     & ...     & ... & 4\\
B2 0034+25 & 128 & 2   & 1210$^{\hskip 0.13cm }$ & 0.72(0.06) & 160(4)   &   0(4) & 67(9)  & 22/17& far     & 64      & 24  & 3,14\\
B2 0915+32 & 248 & 2   & 1540$^{\hskip 0.13cm }$ & 0.28(0.04) & 122(10)  & 22(11) & 88(13) &  1/0 & near    & 80      & 56  & 3,14\\
B2 1256+28 &  90 & 2   &  360$^{\hskip 0.13cm }$ & 0.81(0.10) & 178(3)   &  4(4)  & 43(8)  & 46/43& ...     & ...     & ... & 3,7\\
B2 1339+26 & 303 & 1-2 &  580$^{\hskip 0.13cm }$ & $\sim 0.5$ &   0(10)  & 13(10) & 30(13) & ...  & ...     & ...     & ... & 3\\
B2 1346+26 & 253 & 1   & 6710$^{1}$              & ...        & 134(7)   & 62(11) & 71(11) & ...  & ...     & ...     & ... & 3\\
B2 1357+28 & 251 & 1-2 &  520$^{\hskip 0.13cm }$ & $\ga 0.5$ &  95(5)   &  2(10) & 85(9)  & ...  & ...     & ...     & ... & 3\\
B2 1457+29 & 588 & 1-2 & 5000$^{\hskip 0.13cm }$ & $\ga 0.5$ &  36(4)   & 78(5)  & 61(9)  & ...  & ...     & ...     & ... & 3\\
B2 1525+29 & 261 & 1-2 &  540$^{\hskip 0.13cm }$ & $\ga 0.5$ & 148(5)   & 28(5)  & 62(9)  & ...  & ...     & ...     & ... & 3 \\
B2 2335+26 & 120 & 2   & 1160$^{\hskip 0.13cm }$ & 0.25(0.05) &   6(9)   & 24(9)  & 61(12) & 19/0 & far     & ...     & ... & 3\\
\end{longtable}

\clearpage

\begin{table}
\caption{Host morphologies\label{t:samples}.Relative distribution over host morphology (taken from the LEDA catalogue) for the UGC, UGC~FR-I, UGC non-FR-I samples of galaxies. Col.(2): total number of galaxies in the sample. Col.(3)-(5): the fraction of galaxies within each sample for
the three morphology classes and the Poissonian error.}
\begin{center}
\begin{tabular}{llllll}\hline\hline\noalign{\smallskip}
Sample & N & E & E/SO & SO & comments \\
\hline
UGC total        & 1101 & $0.34 \pm 0.02$ &  $0.16 \pm 0.01$ & $0.50 \pm 0.02$ & $v<7000\kms$ \\
                 &  389 & $0.49 \pm 0.04$ &  $0.17 \pm 0.02$ & $0.34 \pm 0.03$ & $v<7000\kms$ and $M_{\rm B} < -20.4$ \\
UGC~FR-I         &   21 & $0.71 \pm 0.18$ &  $0.19 \pm 0.10$ & $0.10 \pm 0.07$ & \\
UGC non-FR-I     &   52 & $0.71 \pm 0.11$ &  $0.12 \pm 0.05$ & $0.17 \pm 0.06$ & \\
\hline
\end{tabular}
\end{center}
\end{table}

\begin{table}
\caption{UGC non-FR-I sample\label{t:ugcnormal}. General properties of the 52 UGC non-FR-I galaxy sample which forms the 
comparison sample for the UGC~FR-I sample of radio galaxies. Galaxy data
are taken from the LEDA catalogue. Cols.~(1)-(3): galaxy name, Hubble
type and absolute blue magnitude. Cols.(4)-(5): WFPC2
filter name and exposure time. Col.(6): HST program number.}
\begin{center}
\begin{tabular}{ccccrc|ccccrc}\hline\hline\noalign{\smallskip}
NGC & Hubble & $M_B$ & filters & Time & HST &  
NGC & Hubble & $M_B$ & filters & Time & HST  \\
 & Type & (mag) &  & (s) & program & 
 & Type & (mag) &  & (s) & program \\
\hline
507     &  E-S0 &  -22.0 &  F555W & 1600 & 6587 & 4494    &     E &  -21.0 &  F555W & 1000 & 5454 \\
545     &  E-S0 &  -21.4 &  F814W & 1000 & 8683 &         &       &        &  F814W &  460 & 5454 \\
821     &     E &  -20.4 &  F555W &  700 & 6099 & 4526    &    S0 &  -20.8 &  F555W &  520 & 5375 \\
        &       &        &  F814W &  460 & 6099 &         &       &        &  F814W &  520 & 5375 \\
910     &     E &  -21.8 &  F814W & 1000 & 8683 & 4552    &  E-S0 &  -20.9 &  F555W & 2400 & 6099 \\
1016    &     E &  -22.5 &  F555W & 1600 & 6587 &         &       &        &  F814W & 1500 & 6099 \\
1129    &    S0 &  -22.2 &  F555W & 6500 & 6810 & 4589    &     E &  -21.0 &  F555W & 1000 & 5454 \\
        &       &        &  F814W & 6500 & 6810 &         &       &        &  F814W &  460 & 5454 \\
1161    &  E-S0 &  -21.3 &  F547M &  360 & 6837 & 4621    &     E &  -20.7 &  F555W & 1050 & 5512 \\
1497    &    S0 &  -21.0 &  F547M &  300 & 5924 &         &       &        &  F814W & 1050 & 5512 \\
        &       &        &  F791W &  100 & 5924 & 4649    &     E &  -21.5 &  F555W & 2100 & 6286 \\
UGC3426 &    S0 &  -20.7 &  F814W &  260 & 8645 &         &       &        &  F814W & 2500 & 6286 \\
2258    &    S0 &  -21.3 &  F814W & 2700 & 8212 & 4807    &  E-S0 &  -20.7 &  F606W &  400 & 5997 \\
2300    &     E &  -20.9 &  F555W & 1520 & 6099 & 4816    &  E-S0 &  -21.4 &  F606W &  800 & 5997 \\
        &       &        &  F814W & 1450 & 6099 & 4881    &     E &  -20.7 &  F555W & 4000 & 5233 \\
2768    &  E-S0 &  -21.1 &  F555W & 1000 & 6587 &         &       &        &  F814W & 4000 & 5233 \\
        &       &        &  F814W & 2000 & 6587 & 4889    &     E &  -22.6 &  F606W &  320 & 5997 \\
2832    &     E &  -22.4 &  F814W & 2600 & 8184 & 4952    &     E &  -21.6 &  F606W &  400 & 5997 \\
2872    &     E &  -20.5 &  F702W & 1000 & 6357 & 4957    &     E &  -21.2 &  F606W &  400 & 5997 \\
2911    &  E-S0 &  -20.8 &  F547M &  460 & 5924 & 5252    &    S0 &  -21.0 &  F606W &  500 & 5479 \\
3348    &     E &  -21.4 &  F702W & 1000 & 6357 & 5322    &     E &  -21.4 &  F555W & 1000 & 5454 \\
3516    &    S0 &  -20.9 &  F555W & 1000 & 6633 &         &       &        &  F814W &  460 & 5454 \\
        &       &        &  F814W &  730 & 6633 & 5557    &     E &  -21.7 &  F555W & 1000 & 6587 \\
3610    &     E &  -20.8 &  F555W & 1000 & 6587 & 5813    &     E &  -21.0 &  F555W & 1000 & 5454 \\
        &       &        &  F814W & 2000 & 6587 &         &       &        &  F814W &  460 & 5454 \\
3613    &     E &  -20.9 &  F702W & 1000 & 6357 & 5846    &     E &  -21.2 &  F555W & 2200 & 5920 \\
3842    &     E &  -22.2 &  F555W & 1600 & 6587 &         &       &        &  F814W & 2300 & 5920 \\
3894    &  E-S0 &  -21.0 &  F547M &  503 & 5924 & 5982    &     E &  -21.5 &  F555W & 1000 & 5454 \\
4073    &     E &  -22.4 &  F555W & 1600 & 6587 &         &       &        &  F814W &  460 & 5454 \\
4125    &     E &  -21.3 &  F555W & 1000 & 6587 & 6211    &    S0 &  -20.9 &  F606W &  500 & 5479 \\
        &       &        &  F814W & 2000 & 6587 & 6703    &  E-S0 &  -20.8 &  F814W &  320 & 5999 \\
4168    &     E &  -20.7 &  F547M &  460 & 6837 & 7318A   &     E &  -20.9 &  F569W & 1600 & 6596 \\
        &       &        &  F702W & 1000 & 6357 &         &       &        &  F814W & 1000 & 6596 \\
4365    &     E &  -20.9 &  F555W & 1000 & 5454 & 7562    &     E &  -21.1 &  F555W & 2200 & 6554 \\
        &       &        &  F814W &  460 & 5454 &         &       &        &  F814W & 2200 & 6554 \\
4406    &     E &  -20.8 &  F555W & 1000 & 5454 & 7619    &     E &  -21.8 &  F555W & 2200 & 6554 \\
        &       &        &  F814W &  460 & 5454 &         &       &        &  F814W & 2200 & 6554 \\
4472    &     E &  -21.4 &  F555W &  460 & 6673 & 7785    &     E &  -21.3 &  F555W &  800 & 6587 \\
        &       &        &  F814W &  460 & 6673 &         &       &        &        &      &      \\
4473    &     E &  -21.6 &  F555W & 1800 & 6099 &         &       &        &        &      &      \\  
        &       &        &  F814W & 2000 & 6099 &         &       &        &        &      &      \\  
\hline
\end{tabular}
\end{center}
\end{table}

\begin{table}
\caption{Dust properties UGC non-FR-I sample.\label{t:dustugcnormal} Properties of the dust detected in UGC~
non-FR-I sample galaxies. Col.(2): galaxy distance from Faber \etal (\cite{Fab89}) or LEDA, except for NGC
4526, which is taken from Tonry \etal (\cite{Ton01}). Col.(3): dust morphology (see Section~\ref{s:dustproperties} for definitions). 1= dust lane, 2= dust ellipse, 1-2=either lane or ellipse and 3=irregular
dust. Col.(4): longest linear extent of the dust which has a typical relative error of $10\%$.  Entries
marked with a $^1$ have an additional irregular extended component. Col.(5): ellipticity of dust ellipses,
with the measurement error in between brackets. Col.(6): dust position angle and its error in between
brackets. This is the PA of the major axis for dust ellipses and of the longest dust extent for dust lanes.
Col.(7): position angle difference between the galaxy major axis and dust longest axis and its error in
between brackets (see Section~\ref{s:dustproperties}).}
\begin{center}
\begin{tabular}{lccrcrr}\hline\hline\noalign{\smallskip}
NGC & D & morph & size & $\epsilon$ & PA$_{\rm D}$ & $\Delta$PA$_{\rm DG}$ \\
 & (Mpc) &  & (pc) &  & ($\deg$) & ($\deg$) \\
\hline
910     & 69.23 & 1-2 &  160$^{\hskip 0.13cm}$ & $\ga 0.5$ &  ...    & ... \\
1129    & 69.67 & 2   &  670$^{\hskip 0.13cm}$ & 0.85(0.10) &    0(2) &  0(3) \\
1161    & 25.80 & 2   & 2960$^{\hskip 0.13cm}$ & 0.53(0.05) &  135(5) &  2(7) \\
1497    & 81.69 & 3   & 5650$^{\hskip 0.13cm}$ & ...        &  ...    & ... \\
2258    & 53.12 & 3   &  470$^{\hskip 0.13cm}$ & ...        &  ...    & ... \\
2768    & 20.43 & 1   &  120$^{1}$             & ...        &   53(8) & 84(9) \\
2872    & 41.65 & 2   &  150$^{\hskip 0.13cm}$ & 0.46(0.06) &  171(5) &  2(5) \\
2911    & 42.40 & 1   & 1080$^{1}$             & ...        &   63(5) & 85(11)\\
UGC3426 & 53.73 & 3   & 3040$^{1}$             & ...        &  ...    & ... \\
3516    & 34.95 & 3   & 2290$^{\hskip 0.13cm}$ & ...        &  ...    & ... \\
3894    & 42.95 & 1   & 1490$^{1}$             & ...        &  109(4) & 14(4) \\
4125    & 26.48 & 1   &  170$^{1}$             & ...        &  108(10)& 13(18) \\
4406    & 17.77 & 1-2 &   50$^{\hskip 0.13cm}$ & $\la 0.5$ &  ...    & ... \\
4472    & 17.77 & 1   &  210$^{\hskip 0.13cm}$ & ...        &  136(10)& 26(11) \\
4494    & 9.267 & 2   &   70$^{\hskip 0.13cm}$ & 0.50(0.05) &   23(6) &  1(6) \\
4526    & 16.90 & 2   & 2350$^{\hskip 0.13cm}$ & 0.80(0.05) &  127(5) &  5(7) \\
4552    & 17.77 & 1-2 &   40$^{1}$             & $\la 0.5$ &  ...    & ... \\
4589    & 40.40 & 1   & 2310$^{1}$             & ...        &  171(10)& 89(11) \\
4952    & 79.11 & 2   &  190$^{\hskip 0.13cm}$ & 0.54(0.03) &   30(5) &  2(5)  \\
5252    & 89.75 & 3   & 2510$^{\hskip 0.13cm}$ & ...        &  ...    &...    \\
5322    & 22.15 & 2   &  370$^{\hskip 0.13cm}$ & 0.87(0.05) &   88(5) &  2(5) \\
5813    & 31.15 & 3   & 1690$^{\hskip 0.13cm}$ & ...        &  ...    & ... \\
5846    & 31.15 & 3   & 2840$^{\hskip 0.13cm}$ & ...        &  ...    & ... \\
7318A   & 88.80 & 3   & 3790$^{\hskip 0.13cm}$ & ...        &  ...    & ... \\
7785    & 60.05 & 3   & 1610$^{\hskip 0.13cm}$ & ...        &  ...    & ... \\
\hline
%
%
\end{tabular}
\end{center}
\end{table}

\begin{table}
\caption{\label{t:comparedust} Comparison between published and our measurements of
dust ellipticities and PAs. Col.(2)-(3): ellipticity as determined by us with the
error in between brackets and as reported in the literature,
respectively. Col.(4)-(5): position angle of the dust as determined by
us (with error again between the brackets) and as reported in the
literature, respectively. Col.(8): references 1: de Koff \etal (\cite{dKof00});
2: Capetti \& Celotti (\cite{Cap99}); 3: Capetti \etal (\cite{Cap00}); 4: Schmitt
\etal (\cite{Sch02}); 5: van der Marel \& van den Bosch (\cite{vdMar98}); 6: de Ruiter
\etal (\cite{dRui02}); 7: Dufour \etal (\cite{Duf79}). The
reason for the significant difference in the PA$_{\rm D}$ for B2 1346+26, B2 1525+29 as measured by de Ruiter \etal (\cite{dRui02}) and us on the one hand
and by Capetti \etal (\cite{Cap00}) on the other hand is unknown, but the de Ruiter
\etal (\cite{dRui02}) measurements are deemed more accurate (Capetti private comm.).}
%
\begin{center}
\begin{tabular}{lrrrrr}\hline\hline
name & $\epsilon$ &$\epsilon_{\rm ref}$ & PA$_{\rm D}$ & PA$_{\rm ref}$ & ref \\
  &      &   & ($\deg$) & ($\deg$) &  \\
\hline
NGC 383    & 0.23(0.03) & 0.18 & 138(2)  & 135 & 2 \\
NGC 4261   & 0.54(0.03) & 0.58 & 163(1)  & 165 & 2 \\
NGC 7052   & 0.70(0.02) & 0.65 &  65(1)  &  65 & 5 \\
3C449      & 0.54(0.05) & 0.50 & 166(5)  & 169 & 1 \\
NGC 5128   & ...        & ...  & 120(3)  & 122 & 7 \\
3C83.1     & 0.86(0.03) & 0.91 & 168(2)  & 171 & 1 \\
3C296      & 0.71(0.05) & 0.71 & 157(4)  & 160 & 1 \\
3C353      & ...        & ...  & 148(7)  & 166 & 1 \\
3C402N     & ...        & ...  &  58(3)  &  55 & 4 \\
4C-03.43   & ...        & ...  &  86(5)  &  73 & 4 \\
B2 0034+25 & 0.72(0.06) & ...  & 160(4)  & 160/160 & 3/6 \\
B2 0915+32 & 0.28(0.04) & ...  & 122(10) & 125 & 6 \\
B2 1256+28 & 0.81(0.10) & ...  & 178(3)  &   0 & 6 \\
B2 1339+26 & ...        & ...  &   0(10) &   0/175 & 3/6 \\
B2 1346+26 & ...        & ...  & 134(7)  &   0/134 & 3/6 \\
B2 1357+28 & ...        & ...  &  95(5)  &  95/77 & 3/6 \\
B2 1457+29 & ...        & ...  &  36(4)  &  30/50 & 3/6 \\
B2 1525+29 & ...        & ...  & 148(5)  &  25/144 & 3/6 \\
B2 2335+26 & 0.25(0.05) & ...  &   6(9)  &   8 & 6 \\ 
\hline
%
\end{tabular}
\end{center}
\end{table}

\begin{table}
\caption{Fractions of dust morphologies.\label{t:dustmorph} Frequency of dust morphologies as a fraction of total
  number of dust galaxies for the UGC~FR-I and UGC non-FR-I sample.
  Columns (2)-(5) list the fractions of dust lanes, ellipses, dust
  structures which are either ellipses or lanes and irregular dust
  respectively. The definitions of these morphology classes are
given in Section~\ref{s:dustcomparison}.}
\begin{center}
\begin{tabular}{lcccc}\hline\hline
Sample & Lanes & Ellipses & Ambiguous & Irregular \\
\hline
UGC non-FR-I & $0.24 \pm 0.10$ & $0.28 \pm 0.11$ & $0.12 \pm 0.07$ & $0.36 \pm 0.12$ \\
UGC~FR-I     & $0.32 \pm 0.13$ & $0.53 \pm 0.17$ & $0.10 \pm 0.07$ & $0.05 \pm 0.05$ \\
%
\hline
\end{tabular}
\end{center}
\end{table}



\end{document}